\documentclass[twocolumn]{aastex631}

\usepackage{ mathrsfs }
\usepackage{graphicx}
\usepackage{xcolor}
\usepackage{mathtools}
\usepackage{booktabs}
\usepackage{hyperref}
\newcommand{\hst}{\textit{3D-HST}}
\newcommand{\ezp}{\texttt{eazy-py}}
\newcommand{\mstar}{$M_\star$}

\shorttitle{AASTeX v6.3.1 Sample article}
\shortauthors{Gould et al.}

\graphicspath{{./}{Figures/}}

\begin{document}

\title{COSMOS2020: Exploring the dawn of quenching for massive galaxies at $3<z<5$ with a new colour selection method}

\author[0000-0003-4196-5960]{Katriona M. L. Gould}
\affiliation{Cosmic Dawn Center (DAWN)}
\affiliation{Niels Bohr Institute, University of Copenhagen, Jagtvej 128, 2200 Copenhagen N, Denmark}

\author[0000-0003-2680-005X]{Gabriel Brammer}
\affiliation{Cosmic Dawn Center (DAWN)}
\affiliation{Niels Bohr Institute, University of Copenhagen, Jagtvej 128, 2200 Copenhagen N, Denmark}

\author[0000-0001-6477-4011]{Francesco Valentino}
\affiliation{Cosmic Dawn Center (DAWN)}
\affiliation{Niels Bohr Institute, University of Copenhagen, Jagtvej 128, 2200 Copenhagen N, Denmark}

\author[0000-0001-7160-3632]{Katherine E. Whitaker}
\affil{Department of Astronomy, University of Massachusetts, Amherst, MA 01003, USA}
\affil{Cosmic Dawn Center (DAWN)}

\author[0000-0003-1614-196X]{John. R. Weaver}
\affiliation{Department of Astronomy, University of Massachusetts, Amherst, MA 01003, USA}
\affiliation{Cosmic Dawn Center (DAWN)}
\affiliation{Niels Bohr Institute, University of Copenhagen, Jagtvej 128, 2200 Copenhagen N, Denmark}

\author[0000-0003-3021-8564]{Claudia del P. Lagos}
\affiliation{International Centre for Radio Astronomy Research (ICRAR), M468, University of Western Australia, 35 Stirling Hwy, Crawley, WA
6009, Australia}
\affil{ARC of Excellence for All Sky Astrophysics in 3 Dimensions (ASTRO 3D)}
\affiliation{Cosmic Dawn Center (DAWN)}

\author[0000-0001-9705-2461]{Francesca Rizzo}
\affiliation{Cosmic Dawn Center (DAWN)}
\affiliation{Niels Bohr Institute, University of Copenhagen, Jagtvej 128, 2200 Copenhagen N, Denmark}

\author[0000-0002-3560-8599]{Maximilien Franco}
\affiliation{Department of Astronomy, The University of Texas at
Austin, Austin, TX, USA}

\author[0000-0001-5615-4904]{Bau-Ching Hsieh}
\affiliation{Academia Sinica Institute of Astronomy and Astrophysics, No. 1, Sec. 4, Roosevelt Rd., Taipei 10617, Taiwan}

\author[0000-0002-7303-4397]{Olivier Ilbert}
\affiliation{Aix Marseille Univ, CNRS, CNES, LAM, Marseille, France }

\author[0000-0002-8412-7951]{Shuowen Jin}
\affiliation{Cosmic Dawn Center (DAWN)}
\affil{DTU-Space, Technical University of Denmark, Elektrovej 327, DK-2800 Kgs. Lyngby, Denmark}

\author[0000-0002-4872-2294]{Georgios Magdis}
\affiliation{Cosmic Dawn Center (DAWN)}
\affil{DTU-Space, Technical University of Denmark, Elektrovej 327, DK-2800 Kgs. Lyngby, Denmark}
\affiliation{Niels Bohr Institute, University of Copenhagen, Jagtvej 128, 2200 Copenhagen N, Denmark}

\author[0000-0002-9489-7765]{Henry ~J.~McCracken}
\affiliation{Institut d'Astrophysique de Paris, UMR 7095, CNRS, and Sorbonne Universit\'e, 98 bis boulevard Arago, 75014 Paris, France}

\author[0000-0001-5846-4404]{Bahram Mobasher}
\affil{Department of Physics and Astronomy, University of California, Riverside, 900 University Avenue, Riverside, CA 92521, USA}

\author[0000-0002-7087-0701]{Marko Shuntov}
\affiliation{Cosmic Dawn Center (DAWN)}
\affiliation{Niels Bohr Institute, University of Copenhagen, Jagtvej 128, 2200 Copenhagen N, Denmark}

\author[0000-0003-3780-6801]{Charles L. Steinhardt}
\affiliation{Cosmic Dawn Center (DAWN)}
\affiliation{Niels Bohr Institute, University of Copenhagen, Jagtvej 128, 2200 Copenhagen N, Denmark}

\author[0000-0002-6338-7295]{Victoria Strait}
\affiliation{Cosmic Dawn Center (DAWN)}
\affiliation{Niels Bohr Institute, University of Copenhagen, Jagtvej 128, 2200 Copenhagen N, Denmark}

\author[0000-0003-3631-7176]{Sune Toft}
\affiliation{Cosmic Dawn Center (DAWN)}
\affiliation{Niels Bohr Institute, University of Copenhagen, Jagtvej 128, 2200 Copenhagen N, Denmark}
 
\begin{abstract}
We select and characterise a sample of massive (log(M$_{*}/$M$_{\odot})>10.6$) quiescent galaxies (QGs) at $3<z<5$ in the latest COSMOS2020 catalogue. QGs are selected using a new rest-frame colour selection method, based on their probability of belonging to the quiescent group defined by a Gaussian Mixture Model (GMM) trained on rest-frame colours ($NUV-U, U-V, V-J$) of similarly massive galaxies at $2<z<3$. We calculate the quiescent probability threshold above which a galaxy is classified as quiescent using simulated galaxies from the \textsc{shark} semi-analytical model. We find that at $z\geq3$ in \textsc{shark}, the GMM/$NUVU-VJ$ method out-performs classical rest-frame $UVJ$ selection and is a viable alternative. We select galaxies as quiescent based on their probability in COSMOS2020 at $3<z<5$, and compare the selected sample to both $UVJ$ and $NUVrJ$ selected samples. We find that although the new selection matches $UVJ$ and $NUVrJ$ in number, the overlap between colour selections is only $\sim50-80\%$, implying that rest-frame colour commonly used at lower redshifts selections cannot be equivalently used at $z>3$. We compute median rest-frame SEDs for our sample and find the median quiescent galaxy at $3<z<5$ has a strong Balmer/4000 \AA\  break, and residual $NUV$ flux indicating recent quenching. We find the number densities of the entire quiescent population (including post-starbursts) more than doubles from  $3.5\pm2.2\times10^{-6}$ Mpc$^{-3}$ at $4<z<5$ to $1.4\pm0.4\times10^{-5}$ Mpc$^{-3}$ at $3<z<4$, confirming that the onset of massive galaxy quenching occurs as early as $3<z<5$.

\end{abstract}


\section{Introduction} \label{sec:intro}
\noindent In the past few years it has become evident that there exists an extraordinary population of massive galaxies which have already ceased their star formation and become quiescent when the universe was just 2 billion years old ($z\geq3)$. The first hints of this population were candidates selected from optical/near-infrared (NIR) surveys such as in the GOODS-South field \citep{fontana_fraction_2009}, and later the Newfirm Medium-Band Survey (NMBS, \citealp{whitaker_newfirm_2011, marchesini_most_2010}) and the FourStar Galaxy Evolution Survey (ZFOURGE, \citealp{straatman_substantial_2014, spitler_exploring_2014}). Targets were selected using various indicators, either on the basis of their specific star formation rate (sSFR) or rest-frame colours. Although these candidates were tentative then, the recent influx of spectroscopically confirmed massive galaxies at $z>3$ puts aside any doubt that these galaxies were already passive, and rapid evolution in the first few billion years caused the onset of the dawn of quenching. There are several examples of single galaxies (e.g., \citealp{marsan_spectroscopic_2015, forrest_extremely_2020, valentino_quiescent_2020, saracco_rapid_2020, carnall_massive_2023}), as well as full samples \citep{marsan_spectroscopic_2017,schreiber_near_2018,forrest_massive_2020,deugenio_hst_2020, mcconachie_spectroscopic_2021,nanayakkara_population_2022}, that have been spectroscopically confirmed as both massive (usually log(M$_{*}$/M$_{\odot})\gtrsim10$) and quiescent (log(sSFR)\,yr$^{-1}$ $\lesssim-9.5$), as well as a number of studies which selected and analysed massive quenched galaxies in large photometric data sets \citep{merlin_chasing_2018, girelli_massive_2019, merlin_red_2019,guarnieri_candidate_2019, cecchi_quiescent_2019, carnall_timing_2020, shahidi_selection_2020, santini_emergence_2020, stevans_newfirm_2021, ito_cosmos2020_2022}. 

Thanks to the named studies, we now know about the properties of massive quiescent galaxies (QGs) at $z>3$. These galaxies exhibit significant Balmer/$4000$ \AA\ breaks indicative of ages between 100\,Myr and $\sim1$\,Gyr, although the ages can be quite uncertain (e.g. \citealp{schreiber_near_2018, carnall_timing_2020, forrest_extremely_2020, deugenio_hst_2020,nanayakkara_population_2022,carnall_massive_2023}). A few spectroscopically confirmed galaxies exhibit weak flux blue-wards of the break (a ``blue bump"), indicating a stellar population comprising predominantly A-type stars, and therefore likely recent quenching ($<1$Gyr). This is distinct from the UV-upturn observed in local quenched galaxies, which seems to be due to emission from post main sequence stars such as Horizontal Branch (HB) or Asymptotic Giant Branch (ABG) stars (\citealp{greggio_clues_1990,yi_ultraviolet-bright_1997}; see \citealp{dantas_uv_2020} for a more recent overview). The age and mass of such early QGs implies a rapid formation period and recent quenching, much like post-starburst, or E$/$K$+$A galaxies observed at $z<3$ (e.g. \citealp{dressler_spectroscopy_1983, goto_266_2005, wild_post-starburst_2009, ichikawa_recently_2017, wu_fast_2018, chen_post-starburst_2019, wild_star_2020,wu_colors_2020, wilkinson_starburst_2021}; see \citealp{french_evolution_2021} for a review of post-starburst galaxies). An onset of rapid quenching at $z\sim4-6$ as suggested by the previous works would result in a fairly young massive quiescent population at $z\sim3$ which had recently undergone an intense period of star formation. 

Recently quenched QGs at $z>3$ also seem to be fairly dust-free \citep{schreiber_near_2018, kubo_massive_2021}. This could be because dust is being destroyed in the post-starburst phase (e.g., \citealp{akins_quenching_2021}) and our selections for QGs at $z>3$ are dominated by PSBs and relatively dust-free QGs, or because our selections and surveys are blind to dusty QGs at $z>3$ \citep{deugenio_hst_2020}. The upcoming investigations into galaxies appearing in NIR observations with no optical flux (\textit{HST}-dark galaxies) with the \textit{James Webb Space Telescope (JWST)} will hopefully address this pertinent debate \citep{wang_dominant_2019, long_missing_2022,barrufet_quiescent_2021,barrufet_unveiling_2022}. 
  
Quiescent populations are commonly selected using rest-frame colour-colour diagrams, with an increasing number of studies further analyzing the correlations of physical properties relative to the sample location within these diagrams (e.g. \citealp{whitaker_large_2012}, \citealp{belli_mosfire_2019}). The four most popular in use are the $UVJ$ selection criteria (\citealp{wuyts_fireworks_2008}, \citealp{williams_detection_2009}), or variations thereof (e.g. \citealp{whitaker_large_2012}), $BzK$ \citep{daddi_new_2004}, and $NUVrK$ or $NUVrJ$ (\citealp{arnouts_encoding_2013}, \citealp{ilbert_mass_2013}, \citealp{davidzon_cosmos2015_2017}).  Rest-frame colour-colour diagrams have been favoured for their quick and easy use with large photometric data sets; galaxies tend to occupy distinct parameter spaces within these diagrams, making selection for quiescent populations moderately straightforward. Critically, the use of two colours allows for the separation of quiescent from star forming galaxies, with the $U-V$ or $NUV-r$ colour separating star forming galaxies from quiescent, and the $r-J$ or $V-J$ colour separating red dusty galaxies from those that are quiescent. 

These selections were originally designed to separate galaxy populations at low redshift, but as newer data has emerged demonstrating that the bi-modality in populations clearly extends to higher redshift (e.g. \citealp{whitaker_newfirm_2011,muzzin_evolution_2013,ilbert_mass_2013,straatman_substantial_2014}), the selections and variations thereof are now often adopted at $z>2$ (e.g., \citealp{whitaker_large_2012, straatman_substantial_2014, hwang_revisiting_2021, suzuki_low_2022, ji_reconstructing_2022}). In sum, rest-frame colour colour diagrams are ideal to select quiescent populations as they are computationally cheap, straight-forward. They can also be model dependent, but this depends on the way in which the rest-frame colours are calculated. 
 
 However, the application of rest-frame colour diagram selection boxes to higher redshift samples (in particular, $z>3$) should be cautioned. Several studies \citep{leja_beyond_2019-1,belli_mosfire_2019,deugenio_hst_2020} find that $10-30\%$ of $UVJ$ selected QGs at higher redshift are either low redshift interlopers or dusty star-forming galaxies. \citet{lustig_massive_2022} show that this contamination occurs more frequently in the $UVJ$ area populated by older QGs, at the top right corner of the quiescent selection box. Moreover, despite sSFR decreasing with redder $U-V$ and bluer $V-J$ colours, \citealp{leja_beyond_2019-1} demonstrate that a $UVJ$ selection cannot differentiate between moderate and low specific star formation rates. Such a distinction in specific star formation rate is critical at high redshift where recently quenched galaxies, or those in the process of quenching, are more prevalent. Furthermore, \citealp{antwi-danso_beyond_2022} show that the extrapolation of rest frame J (which can be common at $z>3$) leads to galaxies being wrongly classified. 
 
 Whilst rest-frame colour diagram selections remain useful, particularly for large ground based surveys, the current methods explored above are less reliable at $z>3$ for several reasons. Firstly, many studies clearly show that $3<z<5$ is the epoch where QGs first appear (e.g. \citealp{marsan_number_2020},\citealp{valentino_quiescent_2020}), and so the galaxy population is \textit{not yet bimodal}. Instead, as \citealp{marsan_number_2020} find, massive galaxies in this epoch exhibit a wide spectrum of properties, with varying amounts of dust and star formation, as well as some harbouring an Active Galactic Nucleus (AGN) (see also \citealp{marsan_spectroscopic_2015}, \citealp{marsan_spectroscopic_2017} and \citealp{ito_cosmos2020_2022}). It therefore does not make sense anymore to apply a colour colour cut that was designed to separate galaxies in a bimodal universe when that bi-modality does not yet exist. Secondly, it takes on the order of $\sim$Gyrs for galaxies to evolve and become truly red (without dust), which implies that we would not expect to find the upper edge of the $UVJ$ quiescent box highly populated at $z>3$. Instead, it is more likely that galaxies found in this parameter space are instead dusty star forming galaxies that have scattered in over the $V-J<1.5$ line. This is seen clearly in \citealp{lustig_massive_2022}, who compare observations of QGs at $z\sim3$ with several simulations. They find that the contamination of the $UVJ$ quiescent box at $z\sim2.7$ is most significant at reddest edge of the box, near the $V-J<1.5$ line. To remedy this, they suggest changing the $V-J$ constraint from $V-J<1.5$ to $V-J<1$, which effectively cuts the selection area in half.
 
 The third reason is that galaxies in the high redshift universe are faint, and thus even state-of-the-art ground-based surveys will have fairly large photometric errors for most galaxies which propagate to uncertainties a factor of two larger in colours \citep[see Appendix C in][]{whitaker_age_2010}. This is further antagonised by the fact that surveys such as $COSMOS$ are limited to the resolution of the now decommissioned \textit{Infra Red Array Camera} ($IRAC$) on the \textit{Spitzer Space Telescope}, which, besides having a  large point spread function (PSF), also barely probes the Balmer break at $z>4$ for QGs. The use of rest-frame $V-J$ colour as a separator between dusty star forming galaxies and red QGs becomes highly uncertain at $z>5$, where rest-frame $J$ moves beyond $\sim9\mu m$ ($IRAC$/Channel 4) and must therefore be extrapolated. For surveys limited to $\sim5\mu m$ (Channel 2), this becomes $z>3$. All of these reasons together imply that simply drawing a dividing line over a non bimodal, noisy population will result in both incomplete and quite significantly contaminated samples.   
 
 Given the arguments against them, one might reconsider the usefulness of rest-frame colour-colour selections when there are multiple options for spectral energy distribution (SED) fitting tools to calculate sSFRs which can be used to select for QGs. Whilst this alternative is often used (e.g. \citealp{carnall_timing_2020, ito_cosmos2020_2022, carnall_first_2022}), more complex methods may require careful tuning, time, and sometimes vast computing resources. Finally, this still does not solve the problem because these approaches require a sSFR cut or otherwise to select QGs in a population where the lines between star forming and quiescent are blurred. Whilst these methods offer opportunities to study the statistical properties and their stellar assembly, the selection of QGs should still be possible with simpler methods. It is therefore important that we take care in optimizing the rest-frame colour-colour selections as they are fast and can be easily applied to different surveys irrespective of the filters used, therefore serving as crucial tools for the selection of candidates for spectroscopic follow up with instruments such as \textit{JWST}. This is something that the community has already begun to explore, e.g. \citealp{leja_beyond_2019} (efficient selection of QGs at $z<1$) and \citealp{antwi-danso_beyond_2022} (efficient selection of QGs at $z>3$ using synthetic $ugi_{s}$ filters). \\ 
 
In this paper, we present a new rest-frame colour colour diagram and selection method specifically designed to find and select high confidence massive QGs at $z\sim3$ and beyond. Using the latest COSMOS catalogue, COSMOS2020, we explore the utility of this new selection technique. In Sections \ref{data} and \ref{sed}, we describe the data and a modified COSMOS2020 catalogue made specific to this study. In Section \ref{robust} we describe the selection of a robust sample of massive galaxies at $3<z<5$. In Section \ref{qsel}, we introduce the new colour selection method. In Section \ref{results}, we present this new selection applied to COSMOS2020 and the main results. Finally, we summarise our conclusions and outlook in Section \ref{con}. For all calculations we use the WMAP9 flat LambdaCDM cosmology \citep{hinshaw_nine-year_2013} with $H_{0} = 69.3\,\mathrm{km\,s^{-1}\,Mpc^{-1}}$, $\Omega_{\rm m}=0.307$. All magnitudes are in the AB system defined by \citealp{oke_secondary_1983} as $m_{AB} = -2.5$log$_{10}\big(\frac{f_{\nu}}{3631Jy}\big)$.

\section{Data}\label{data}
\subsection{COSMOS2020}
\noindent The Cosmological Evolution Survey ($COSMOS$) is currently the deepest NIR wide-field multi-wavelength survey that provides data over 2 square degrees of the sky (\citealp{scoville_cosmic_2007}, \citealp{koekemoer_cosmos_2007}). The latest version of the $COSMOS$ photometric catalogue, $COSMOS2020$ \citep{weaver_cosmos2020_2021}, improves on previous version \citep{laigle_cosmos2015_2016} in many ways. Firstly, with the addition of ultra-deep optical data from \textit{Hyper-Suprime Cam} ($HSC$; \citealp{miyazaki_hyper_2018, aihara_hyper_2018}), secondly, the fourth data release of \textit{Ultra-VISTA} (DR4) \citep{mccracken_ultravista_2012,moneti_fourth_2019}, effectively 1\,mag deeper in $K_{s}$ over the entire field, and thirdly, the inclusion of all \textit{Spitzer}/$IRAC$ data ever taken in the $COSMOS$ field \citep{weaver_cosmos2020_2021,moneti_euclid_2022}. $COSMOS2020$ contains four catalogues, combining two source extraction methods and two photometric redshift codes. The Classic catalogue is produced using classical aperture based photometry method using \textsc{Source Extractor} \citep{bertin_sextractor_1996}, whilst \textsc{The Farmer} catalogue is produced using a profile fitting photometry method (Weaver at al., submitted), which is a wrapper written for the source profile fitting code \textsc{The Tractor} \citep{lang_tractor_2016}. \textsc{The Farmer} has a distinct advantage over classical photometry methods due to its ability to correctly model and de-blend overlapping sources. This is particularly beneficial in deep fields where source crowding and overlapping is common. Both photometry catalogs are fit with two different photometric redshift codes (Le Phare, \citealt{arnouts_lephare_2011}, and \ezp\, \citep{brammer_eazy_2008, brammer_eazy-py_2021}. \citealp{weaver_cosmos2020_2021} show that \textsc{The Farmer} performs equivalently to \textsc{Source Extractor} and excels at $i<24$, independent of the redshift code used. For this analysis we use \textsc{The Farmer} photometric catalogue combined with redshift and physical parameter estimation using \ezp.\ 

Although \textsc{the Farmer} photometry and its associated uncertainties are suitable for general use SED fitting, \citet{weaver_cosmos2020_2021} noted that it is likely the photometric uncertainties are underestimated. To test this, we ran the custom code \textsc{Golfir} \citep{kokorev_alma_2022} on a small subset of the COSMOS2020 $IRAC$ images. \textsc{Golfir} uses a prior based on the highest available resolution imaging to model the flux in the $IRAC$ photometry. Overall, the agreement of both photometric catalogs is high, however the $IRAC$ errors derived with \textsc{The Farmer} are smaller by a factor of $\sim5$. This is similar to the approach and conclusions in \citet{valentino_archival_2022}. At $z>3$, the $IRAC$ photometry plays a crucial role in the identification of massive galaxies. This is due to the fact that the light emitted from the bulk of the stellar mass is measured at observed frame $\lambda_{obs} > 2.5 \mu{\rm m}$, which is measured predominantly by the $IRAC$ bands. Although the depths probed by the first two $IRAC$ bands Channels 1 and 2 are comparable to those in \textit{Ultra-VISTA} bands $JHK_{s}$ ($\sim 25.5$ mag, $3\sigma$), $IRAC$ channels 3 and 4 are shallower by $\sim 3$ mags. For massive QGs at $z>3$, this may result in the $IRAC$ bands approaching the detection limit of the survey. For this reason and the investigations explained above, for this work we conservatively boost the IRAC photometric errors by a factor of $5$ and refit the $COSMOS2020$ \textsc{Farmer} photometry with \ezp\ (see Section \ref{sed}). As with \citet{weaver_cosmos2020_2021}, we do not include the $Subaru$ \textit{Suprime-Cam} broad band photometry or the $GALEX$ $FUV$ and $NUV$ photometry, due to their shallow depth and broad PSF, respectively. In total, we use 27 photometric bands for SED fitting, probing observed frame $\sim 0.3-8 \mu{\rm m}$. 

\subsection{3D-HST}\label{3dhst}
In Section \ref{sed}, we will introduce the latest python implementation of \texttt{eazy}, \ezp\, and its newer features. To demonstrate their validity, in particular the computation of physical properties such as stellar mass and star formation rate, we use an observational catalog with similar properties as $COSMOS2020$ to benchmark the performance. We consider the \textsc{Prospector} version of the \hst\ catalog from \citep{leja_older_2019}. The \hst\  survey \citep{skelton_3d-hst_2014} is a $248-$orbit Hubble Treasury program completed in 2015, providing $WFC3$ and $ACS$ spectroscopy for galaxies in 5 extra-galactic deep fields, including $COSMOS$, covering 900 square arc-minutes (199 square arc-minutes in $COSMOS$). The catalog includes photometry in up to 44 bands in $COSMOS$ covering 0.3-8\micron\ in the observed frame, with the inclusion of \textit{Spitzer} $MIPS$ 24\micron\ photometry from \citet{whitaker_constraining_2014}. The version that we considered contains a representative sub sample of $58,461$ galaxies from the \hst\ survey (roughly 33\%) which span the observed star formation rate density and mass density at $0.5<z<2.5$. These galaxies are fit with the Bayesian SED fitting code \textsc{Prospector} \citep{leja_deriving_2017,johnson_stellar_2020}. \textsc{Prospector} uses galaxy models generated using \textsc{python-FSPS} \citep{conroy_propagation_2009,conroy_propagation_2010}, non parametric star formation histories, variable stellar metallicities and a two-component dust attenuation model, under the assumption of energy balance. For this paper, we cross match our catalog and the \hst\ \textsc{Prospector} catalog using a $0.\arcsec6$ matching radius, resulting in 8964 galaxies in common.

\subsection{Simulated data}
\label{sec:shark}
To test the robustness of our quiescent galaxy selection method, we use simulated galaxies from the {\sc Shark} semi-analytical model \citep[SAM,][]{lagos_shark_2018}. {\sc Shark} models the formation and evolution of galaxies following many physical processes, including star formation, chemical enrichment, gas cooling, feedback from stars and active galactic nuclei, among others. Critically for this work, {\sc Shark} includes a two-phase dust model for light attenuation that is based on radiative transfer results from hydrodynamical simulations \citep{trayford_fade_2020}. This model was combined with the stellar population synthesis model of \citet{bruzual_stellar_2003}, adopting a \citet{chabrier_galactic_2003} initial mass function, and a model for re-emission in the IR in \citet{lagos_far-ultraviolet_2019}. {\sc Shark} does this using the generative SED mode of the {\sc ProSpect} SED code of \citet{robotham_prospect_2020}. The predicted SED of each galaxy is then convolved with a series of filters to get their broad-band photometry. \citet{lagos_far-ultraviolet_2019} showed that {\sc Shark} is very successful in reproducing the FUV-to-FIR emission of galaxies in observations at a wide redshift range, including luminosity functions, number counts and cosmic SED emission. \citet{lagos_physical_2020} used this model to show that the contamination fraction of dusty, star-forming galaxies (those with $870\,\mu{\rm m}$ flux $>0.1$~mJy) in the $UVJ$ diagram increases with increasing redshift, reaching $\approx 47$\% at $z=3$, showing that the commonly adopted $UVJ$ colour selection could be improved to reduce contamination. Here, we use a lightcone presented in \citet{lagos_far-ultraviolet_2019} (see their Section~5) that has an area of $\approx 107\,\rm deg^2$, and includes all galaxies in the redshift range $0\le z \le 8$ that have a dummy magnitude, computed assuming a stellar mass-to-light ratio of $1$, $<32$. The method used to construct lightcones is described in \citet{chauhan_h_2019}.

\section{SED fitting with eazy-py}\label{sed} 

\subsection{Eazy-Py}
In this study, we use photometric redshifts, stellar masses, star formation rates and rest-frame colours derived with \ezp\footnote{https://github.com/gbrammer/eazy-py} \citep{brammer_eazy-py_2021}. Here we detail this process, including an updated description and presentation of the \ezp\ code. \ezp\ is a pythonic photometric redshift code adapted from the original \texttt{eazy} code written in C$+$ \citep{brammer_eazy_2008}. \ezp\ estimates redshifts by fitting non-negative linear combinations of galaxy templates to broad-band photometry to produce a best fit model. The flexibility of the code means that in theory, any kind of template can be used, e.g., quasar or AGN templates. The templates used in this paper are comprised of 13 stellar population synthesis templates from FSPS \citep{conroy_propagation_2009,conroy_propagation_2010} spanning a wide range in age, dust attenuation and log-normal star formation histories.  This template set is designed to encompass the expected distribution of galaxies in $UVJ$ color space at $0<z<3$, and in theory beyond. We use a Chabrier IMF \citep{chabrier_galactic_2003} and the \citealp{kriek_dust_2013} two component dust attenuation law, which allows for a variable UV slope (dust index $\delta = -0.1$, RV $= 4.05$). 

\begin{figure}
    \centering
    \includegraphics[width=0.4\textwidth]{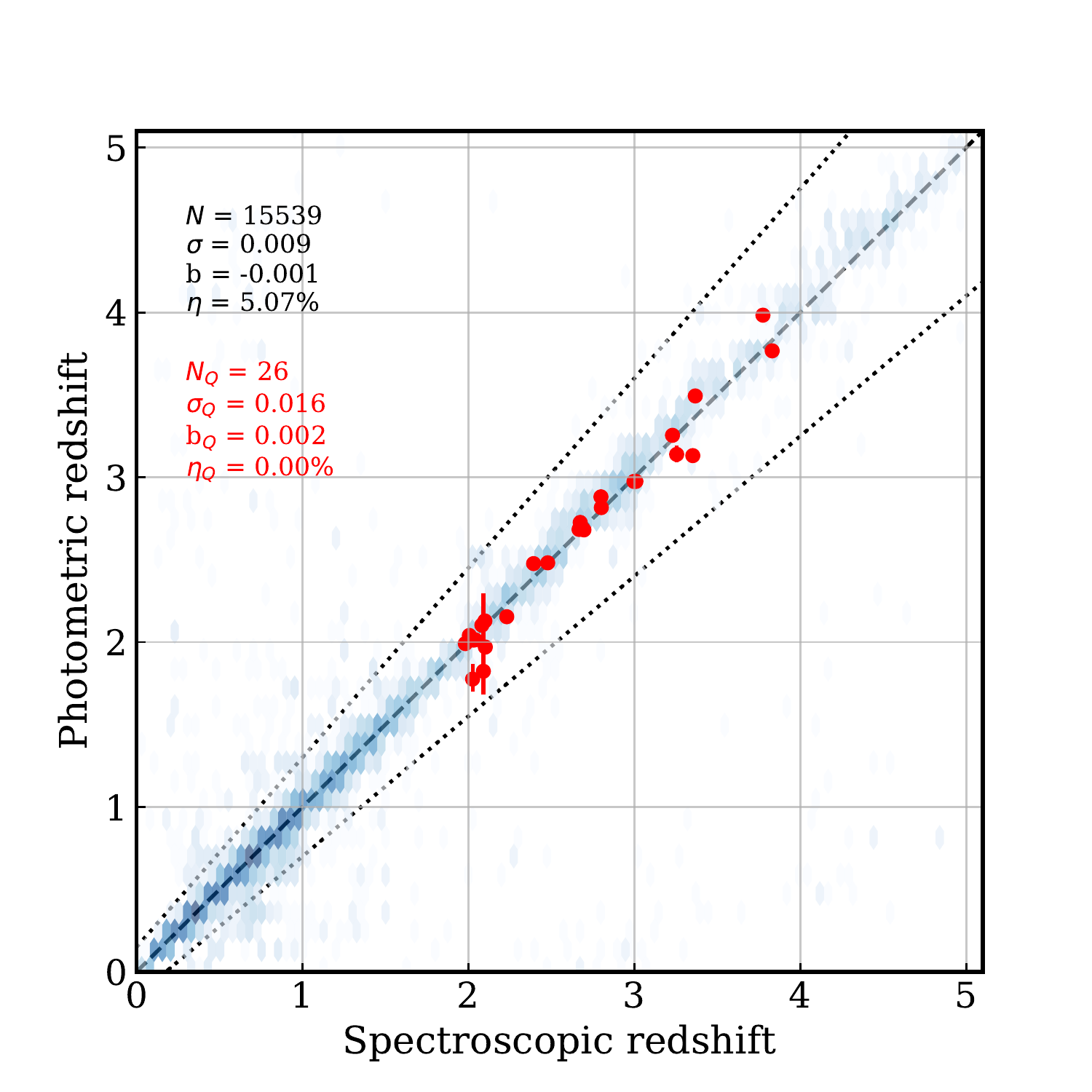}
    \caption{Photometric redshift vs spectroscopic redshift for all galaxies at $0<z<5$ in the combined UVISTA and HSC area with $K_{s}$ SNR$>5$. Red points are spectroscopically confirmed QGs at $z>2$ from \citealp{stockmann_x-shooter_2019}, \citealp{deugenio_hst_2020}, \citealp{schreiber_near_2018}, \citealp{forrest_massive_2020}, \citealp{valentino_quiescent_2020}.} 
    \label{fig:zz}
\end{figure}

The latest version of \ezp\ (version 0.5.2) has been adapted to be a general use SED fitting code, providing estimates of physical parameters such as stellar mass, star formation rate, and dust attenuation. These physical properties are assigned to each template and ``propagate through" the fitting process with the same non-negative linear combination coefficients that return the best-fit SED model. This results in star formation histories that are non-negative linear combinations of a number of basis star formation histories, and therefore the star formation history of a galaxy best-fit model will mimic non-parametric methods such as those used in \citealp{pacifici_evolution_2016, iyer_nonparametric_2019, leja_how_2019} and other works. This means that \ezp\ is more similar to codes that implement non-parametric star formation histories than it is to traditional SED fitting codes, and can account for the mass that is often missed as a result of using parametric star formation histories \citep{lower_how_2020}. We demonstrate this in Section \ref{masses}. Errors on physical parameters are calculated by drawing 100 fits from the best fit template error function and using the $16$th and $84$th percentiles as the extremes of the 1$\sigma$ errors. It should be noted that the physical parameter errors are not marginalised over the redshift error, therefore it is likely that the errors from \ezp\ are underestimated. Figure \ref{fig:zz} shows a comparison of photometric and spectroscopic redshifts for all galaxies at $0<z<6$ in our catalog. We find similar outlier fractions and bias for both star forming and QGs compared to the official COSMOS2020 release \citep{weaver_cosmos2020_2021}. 

\subsubsection{Rest-frame flux densities}\label{rest}

The rest-frame fluxes are calculated based on the approach of \citet{brammer_number_2011}, which uses the best fit model as a guide to interpolate between the observed photometry. This interpolation itself is weighted by the photometric errors, and therefore the rest-frame flux errors reflect those of the photometry. This means that the rest-frame fluxes are almost entirely derived from the photometry with only partial guidance from the best fit template. The results also account for the filter shapes and relative depths, which is advantageous particularly for multi-instrument photometric catalogs. This essentially ensures a ``model-free'' approach, which is crucial for our method, as rest frame colours are used in the selection process, and should ideally reflect the \textit{observed} universe and not our models. For a full description of calculation of the rest-frame fluxes we refer the reader to Appendix \ref{rf-fluxes}. The rest-frame filters used in this paper are the GALEX $NUV$ band ($\lambda = 2800$\AA) \citep{martin_galaxy_2005}, the re-calibrated  $U$ and $V$ filters by \citet{maiz_apellaniz_recalibration_2006}, and 2MASS $J$ band \citep{skrutskie_two_2006}.

\subsubsection{Stellar masses and Star formation rates}\label{masses}

\begin{figure*}
  \centering
  \begin{minipage}[b]{0.49\textwidth}
    \includegraphics[width=\textwidth]{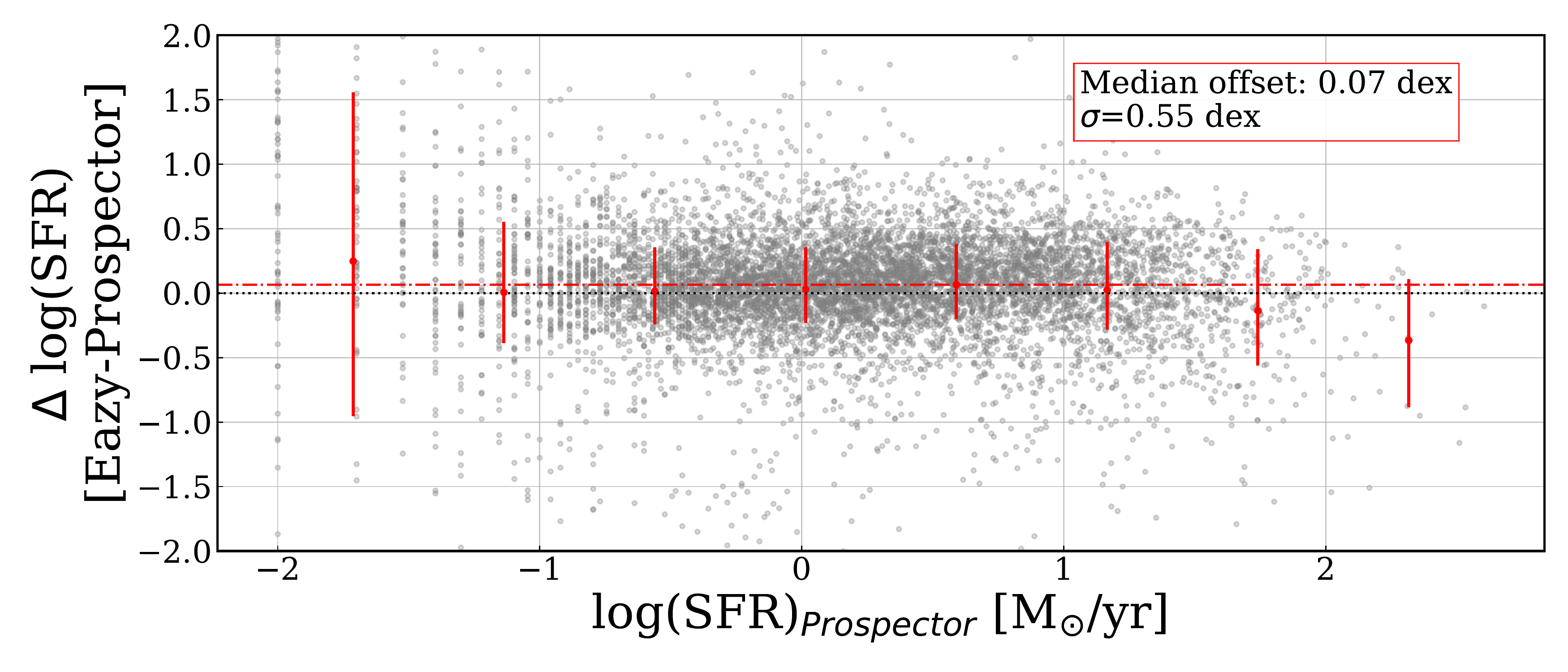}
  \end{minipage}
  \hfill
  \begin{minipage}[b]{0.49\textwidth}
    \includegraphics[width=\textwidth]{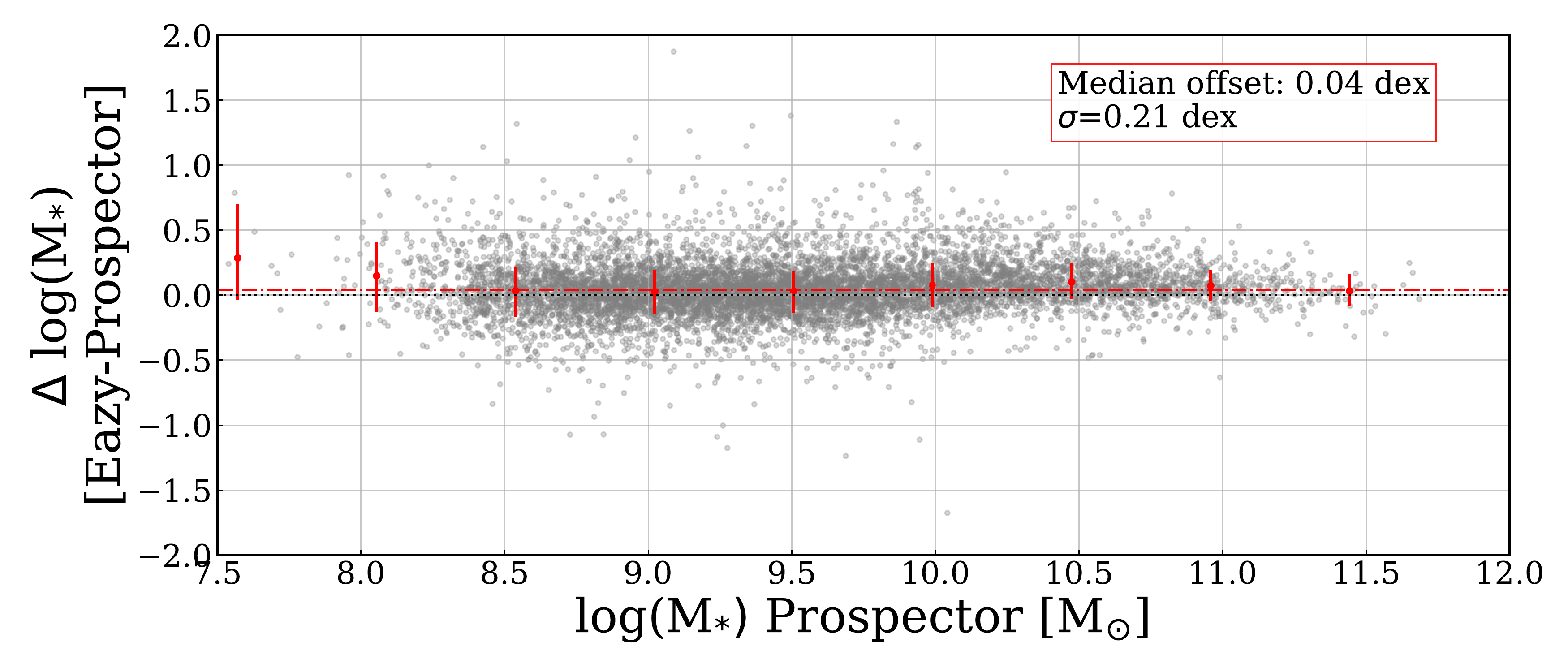}
  \end{minipage}
  \caption{\textit{Left}: $\Delta$log(SFR) (\textsc{Eazy}$-$\textsc{Prospector}) as a function of \textsc{Prospector} SFR for 8134 galaxies at $0.5<z<3.0$ matched to our catalogue with that of the $3D$-HST survey in COSMOS. The median $\Delta$log(SFR) is 0.07 dex (shown as red dashed line) and the scatter is 0.55 dex. The binned median offset and the associated 16th and 84th percentiles are shown as red points with error bars. \textit{Right}: $\Delta$log(M$_{*}$) (\textsc{Eazy}$-$\textsc{Prospector}) as a function of \textsc{Prospector} stellar mass for the same sample. The median $\Delta$log(M$_{*}$) is 0.04 dex and the scatter is 0.21 dex. The binned mean offset and the associated 16th and 84th percentiles are shown as red points with error bars.}
  \label{compare}
\end{figure*}

In this section we explore the stellar masses ($M_\star$) and star formation rates (SFRs) derived with \ezp.\ We note that it has been shown that star formation rates derived from broad-band photometric SED fitting assuming parametric star formation histories can underestimate the SFR of a galaxy by several orders of magnitude \citep{lower_how_2020}. \citet{sherman_investigating_2020}, who used \ezp\ with default FSPS templates using a variety of realistic SFHs (rising, bursty), find that $M_\star$ are overestimated on average by $0.085$ dex, and SFRs are underestimated on average by $\sim0.46$ dex. As we fit with a different, newer set of templates using log-normal star formation histories, we cannot assume the derived masses and SFRs have from the same offsets. 

To benchmark the \ezp\ $M_\star$ and SFRs in this study, we use the $3D$-HST catalog described in Section \ref{3dhst}. From the \textsc{prospector} 3D$-HST$ catalog we use redshift (which are a compilation of spectroscopic redshifts, grism redshifts and redshifts derived using \texttt{eazy}), stellar masses and star formation rates, which are averaged over the past 100 million years. Here, we use the masses and star formation rates derived with Prospector as the ``ground truth". We select all galaxies with a photometric redshift agreeing within $15\%$ of $\Delta z/(1+z)$ to the \textsc{Prospector} redshift, which results in 8134 galaxies spanning $0.5<z<3.0$. We then compute the difference (in dex) between our parameters (SFR, $M_\star$) and the \textsc{Prospector} parameters. Due to differences in method, the $K_{s}$ band magnitudes for COSMOS2020 are marginally brighter ( $\sim0.09$ dex) than those in the \citealp{skelton_3d-hst_2014} catalog that was used in \citealp{leja_how_2019}. To facilitate a fairer comparison, we scale the Prospector masses and star formation rates accordingly. In practice, this results in a very small difference. We show both the SFR and mass comparisons in Figure \ref{compare} as a function of the estimates derived with \textsc{Prospector}. We calculate the median offset and scatter, and find that there is excellent agreement between the two catalogs, with only a minor overestimation of 0.04 dex for masses and 0.07 dex for SFRs by \ezp\, with a scatter of $0.21$ dex for mass and $0.55$ dex for SFRs. The difference between the results of this investigation compared to \citealp{sherman_investigating_2020} is notably that of the SFRs - whilst they find an underestimation of $\sim$0.5 dex, we report a slight over-estimation. This is likely due to the different template sets used. In conclusion, given the agreement and no catastrophic bias in stellar masses or SFRs derived with \ezp\, we deem them acceptable for use.

\section{Selecting a robust sample of massive galaxies}\label{robust}

We begin by selecting all objects within the area covered by both $HSC$ and \textit{Ultra-VISTA} (i.e., \textsc{Flag Combined} $=0$, area $\approx$ 1.27$^{\circ 2}$), classified as galaxies according to the star galaxy separation criteria (see Appendix \ref{sgs}), and with $K_s < 25.6$ mag, which is the $3 \sigma $ limiting magnitude. From these, we select all galaxies with photometric redshifts $3.0<z_\mathrm{phot}<5.0$, with a constraint that the lower $16\%$ of the photometric redshift probability distribution, $p(z)$, should be contained above $z=3$, and the upper $84\%$ below $z=5$. If a spectroscopic redshift is available in the catalogue we require $3<z_\mathrm{spec}<5$. At $4.5<z<5$, the rest-frame $J$ band is covered by IRAC channels 3 and 4, which means that it is still observationally constrained and not extrapolated (see Appendix \ref{app:robustness_jband}). This is vital to ensure the robustness of the sample (\citealp{antwi-danso_beyond_2022}). For this reason, we do not extend beyond $z=5.0$ in our sample selection. We next make a cut on stellar masses, 10.6 $<$ log(M$_{*}/$M$_{\odot}) < 12$, which is the knee of the mass function at this redshift range and well above the stellar mass completeness \citep{weaver_cosmos2020_2021}. However, it is important to note that at $4.5<z<5$, the Balmer/$4000\AA$ break moves through the $K_{s}$ band, which is the reddest band in the chi-mean $izYJHK_{s}$ detection image used in the photometric modelling. Therefore, our selection is likely not sensitive to older quiescent galaxies at this redshift range, if they exist. To our sample, we then add the following requirements to ensure the confidence in our massive galaxy sample is high. We choose to conservatively exclude anything with a reduced $\chi^{2}$ equal to $\chi^{2} / (N_\mathrm{filt})$ greater than 10. We also require  K$_{s}$ SNR $>5$. In summary, we require:

\begin{itemize}
    \item Inside combined HSC/UVISTA area
    \item Classified as galaxy 
    \item K$_{s}$ MAG$<25.6$
    \item $3.0<z_{phot}<5.0$ (or $3.0<z_{spec}<5.0$ if available)
    \item $z_{phot, 16\%}>3.0$ and $z_{phot, 84\%}<5.0$
    \item $10.6$ $<$ log(M$_{*}/$M$_{\odot})<12$
    \item Reduced $\chi^{2}<10$
    \item K$_{s}$ SNR $>5$
\end{itemize}

\noindent In total, this selection results in 1568 galaxies with a median redshift of $z_{phot}=3.55$ and median stellar mass of log(M$_{*}/$M$_{\odot})=10.94$. In addition to this, we select a sample at $2<z<3$ in a similar manner with a lower mass cut of $10$ $<$ log(M$_{*}/$M$_{\odot})<12$, in order to fit the Gaussian Mixture Model described in the following section. This gives a fitting sample of 12985 galaxies at $2<z<3$ with a median photometric redshift of $z_{phot}=2.46$ and median stellar mass of log(M$_{*}/$M$_{\odot})=10.52$. 

\section{Quiescent galaxy selection method}\label{qsel}

\subsection{Introducing the GMM} \label{newcolor}

\begin{figure*}
    \centering
    \includegraphics[width=0.99\textwidth]{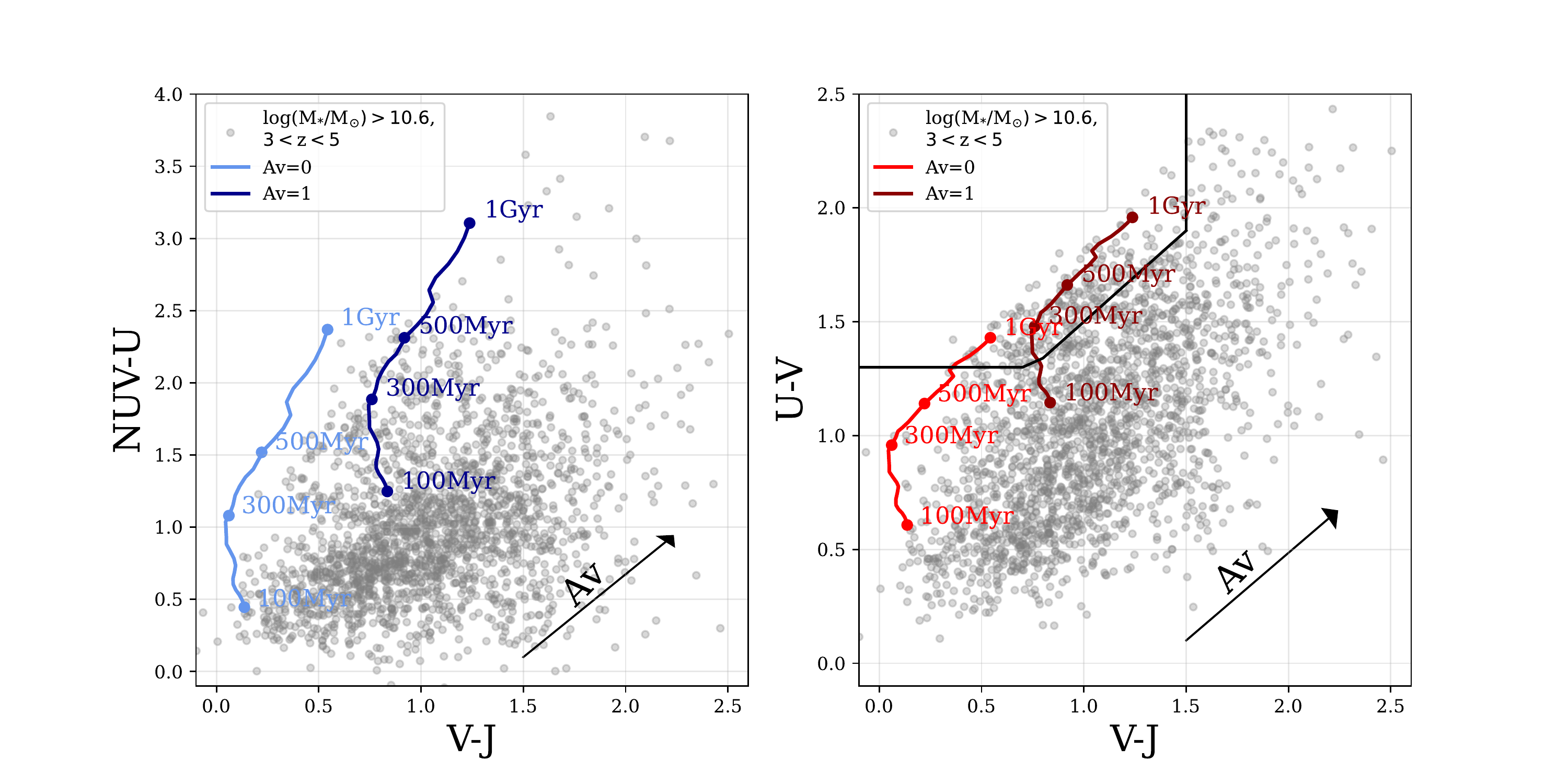}
    \caption{\textit{Left}: $NUV-U$, $V-J$ colours of our massive ($10.6$ $<$ log(M$_{*}/$M$_{\odot})<12$) galaxy sample at $3<z<5$ (grey) and evolution tracks for an SSP with $A_{\rm V}=0$ (blue line on the left of the figure) and $A_{\rm V}=1$ (dark blue line) and Chabrier IMF aged to 1 Gyr. The dust vector representing the movement of a galaxy in this space by attenuation of $A_{\rm V}=1$ is shown by the arrow at the bottom right. \textit{Right}: Same but for $UVJ$ with the quiescent box defined by \cite{whitaker_large_2012} (defined as $U - V > 0.8 \times (V - J) + 0.7$, $U - V > 1.3$ and $V - J < 1.5$).}
    \label{fig:tracks}
\end{figure*}

\noindent We present here a new colour selection designed to make finding QGs at $z>3$ easier. Our new colour selection method combines the $NUV-U$, $U-V$ and $V-J$ colours, using the filters defined in Section \ref{rest}. This colour selection method is \textit{similar} to both \textit{$UVJ$} and \textit{NUVrJ}, with two major differences. The first is that four bands (three colours) are used instead of three, and the second is that the separation itself is not defined by a traditional ``box" but rather using simple machine learning methods. Both are introduced to make the separation of QGs from (dusty) star forming more pronounced, and furthermore allow an easier separation between recently quenched and older QGs. The $U-V$ and $V-J$ colours are included to separate quiescent galaxies from star forming, as originally designed. The $NUV-U$ colour is added to measure the light emitted from recent star formation in the form of A-type stars and is similar to $NUV-r$, except that it probes a shorter baseline, specifically the flux directly blue-wards of the 4000 \AA/Balmer break. Previous works have suggested $NUV-U$ is a viable indicator for time since quenching, for example \cite{schawinski_green_2014}. Similarly, \cite{phillipps_galaxy_2020} use $NUV-u$ as a means to remove passive galaxies with some residual star formation from a sample of truly passive galaxies. The reason for the use of four bands / three colours is twofold: firstly, the increase in dimensions allows us to extract more information whilst still only requiring 4 data points. Whilst this is similar to dimensionality reduction, it has the advantage of using information that is likely already available or easy to compute, and does not require imposing a high signal to noise cut to the data. Secondly, $NUV-U$ does not have a co-dependent variable with $V-J$, which allows for the spread of colours in $NUV-U$/$V-J$ to become more obvious, making it easier to both separate the quenched population from star forming, as well as explore the properties of galaxies within the quenched population with different amounts of recent star formation. This is particularly relevant at $z>3$, where the fraction of PSB galaxies is high (e.g. \citealp{deugenio_typical_2020} and \citealp{lustig_compact_2020} report a PSB fraction of 60-70$\%$ for photometrically selected QGs at log(M$_{*}/$M$_{\odot})>11$). If we consider the 2D projection ($NUV-U$, $V-J$), it is clear that the star forming and quiescent populations separate more clearly due to the divergent alignment of the dust vector relative to the evolution tracks (see Figure \ref{fig:tracks}). As an example, we plot the massive $3<z<5$ sample in this colour space with tracks showing the colour evolution of a single stellar population aged to 1 Gyr at both $A_{\rm V}=0$ and $A_{\rm V}=1$ generated using the Python-FSPS package. Both tracks move directly up through the quiescent area as they evolve over time. The lower portion of this area preferentially selects quenching or post-starburst galaxies, whilst the upper half encompasses older QGs. This age trend has been observed in the $UVJ$ diagram (e.g. \citealp{whitaker_large_2012}, \citealp{whitaker_quiescent_2013}, \citealp{belli_mosfire_2019}, \citealp{carnall_how_2019}), thus it is not entirely surprising that it appears in this colour diagram too -- and in fact more pronounced due to the addition of the $NUV-U$ colour.  We note that the direction of the age tracks diverge from the direction of the dust vector for ($NUV-U$, $V-J$), resulting in red dusty star forming galaxies pushed downwards and to the left relative to old, red QGs, whereas for $UVJ$ they follow the same direction, resulting in old, red QGs populating the same space as red dusty star forming galaxies.

\subsection{Defining the separator} \label{sepsection} 

\begin{figure*}[!tbp]
  \centering
  \begin{minipage}[b]{0.99\textwidth}
    \includegraphics[width=\textwidth]{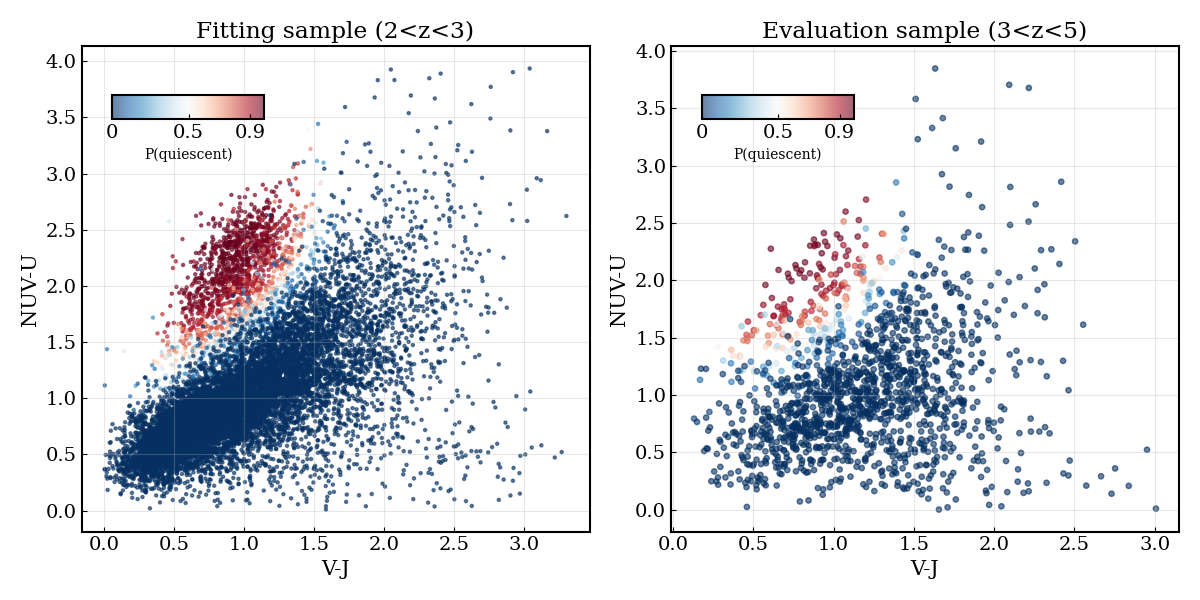}
  \end{minipage}
  \caption{\textit{Left:} $NUV-VJ$ diagram with the $2<z<3$ fitting sample coloured by $p(q)_{50}\%$, which is the median of the boot-strapped quiescent probability distribution based on the Gaussian Mixture Model. \textit{Right}: Same but for the $3<z<5$ sample.} 
  \label{pq}
\end{figure*}

\noindent Most selections in rest-frame colour diagrams have been made by drawing a box empirically. We choose to separate QGs from star forming by using a Gaussian Mixture Model (GMM). This is a probabilistic model that operates under the assumption that the data can be fit by a finite number of Gaussian curves, for which the parameters are not known. GMMs have already been employed successfully in multiple research areas in astrophysics, including to select and explore various galaxy populations such as quiescent and PSB galaxies (e.g. \citealp{black_red_2022,ardila_shocked_2018}). Probabilistic selection of QGs at $z>3$ has also been carried out with great success by \cite{shahidi_selection_2020} and \cite{santini_emergence_2020}. We use the Gaussian Mixture Model package supplied by Scikit-learn \citep{pedregosa_scikit-learn_2011}. Briefly, the GMM algorithm uses expectation-maximisation (EM), which is an iterative process used for classification when there are no ``correct" labels (as is our case). EM works by choosing random components and iterating towards the best fit by computing the likelihood of each data point being drawn from the current model, and then adjusting the parameters to maximise that likelihood. In this approach, we choose to use all three colours: $NUV-U$, $U-V$ and $V-J$. We fit the model using the $2<z<3$ sample, which is first cleaned by requiring all colours have $0<colour<4$. To ascertain the optimal number of components, we fit multiple times using $1-10$ Gaussian components and calculate the Bayesian Information Criterion (BIC) for each number of components, which is defined as 

\begin{equation}
    BIC = k\text{ln}(n) - 2\text{ln}(\mathscr{L}_{\text{max}}) 
\end{equation}

\noindent where $n$ is the number of model parameters, $k$ is the number of data points fit and $\mathscr{L}_{\text{max}}$ is the maximum likelihood of the fit. The BIC discriminates between models by penalising the number of parameters, which avoids over-fitting. We randomly select $60\%$ of the sample, fit and compute the BIC for $1-10$ components, and repeat 1000 times. We find that a $6-$component model is best fit in $\sim75\%$ of the repeated fits and adopt this as our baseline model. The GMM returns for each galaxy a likelihood of belonging to each group, which we convert to a probability. We could simply use this value to decide what to classify as quiescent, however, this does not account for the errors in a galaxy's colours. To assign each galaxy a probability of belonging to the quiescent group, which is identified by eye as the component in the quiescent area of $UVJ$ space, we instead take the following approach, inspired by \cite{hwang_revisiting_2021}. Each galaxy has rest frame fluxes in the $NUV$, $U$, $V$ and $J$ bands, and associated errors. Assuming these are Gaussian distributed, we boot-strap the fluxes 1000 times and compute the $NUV-U$, $V-J$ and $U-V$ colours for each set of rest frame fluxes. Each set of 1000 colours is then fit with the GMM and the percentiles ($5,16,50,84,95$) of the probabilities are calculated. This results in a quiescent probability distribution $p(Q)$ \textit{for each galaxy}, which is marginalised over its rest frame flux errors. Therefore, the final distribution of $p(Q)$ for each galaxy accounts for flux errors. We refer mainly to $p(q)_{50}$ (abbrev. $p(q)$) throughout the text, which is the median quiescent probability calculated in this way. Here, we use the GMM purely as a tool to classify QGs and do not explore the other five groups defined by the model. In Figure \ref{pq} we show the $2<z<3$ fitting sample and the $3<z<5$ sample in the $NUVU-VJ$ plane, coloured by their quiescent probability $p(q)$. It is evident that the model correctly finds the boundary between quiescent and star forming, but a boundary that is a smooth gradient rather than a binary separation. 

\subsubsection{The $p(q)$ threshold for quiescence}

\begin{figure*}
    \centering
    \includegraphics[width=\textwidth]{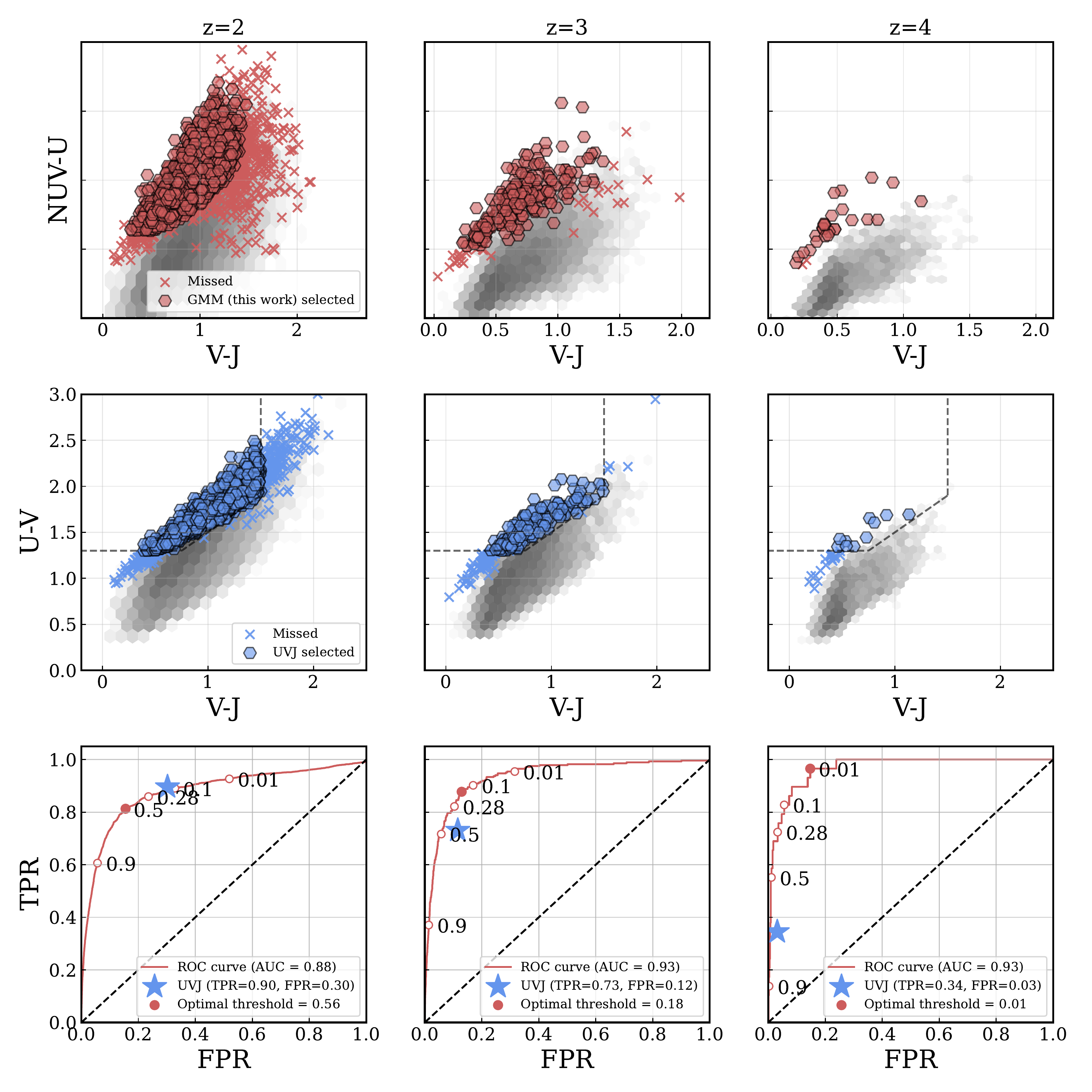}
    \caption{\textit{Left column}: \textsc{SHARK} simulated galaxies in different colour planes. GMM/$NUV-VJ$ (this work, top) and $UVJ$ (middle) colour diagrams for massive (log(M$_{*}/$M$_{\odot}>10$) galaxies in the {\sc Shark} simulation (grey), with QGs found by each selection method as red (blue) hexagons and galaxies missed by both selection methods as red (blue) crosses. The bottom panel shows the Receiver Operator Characteristic (ROC) curve for the trained Gaussian Mixture Model as a function of quiescent probability ($p(Q)$) threshold. The Area Under the Curve (AUC), which represents the percentage probability of correctly classifying QGs, is reported, as well as the $p(Q)$ threshold, which is calculated as the geometric mean (shown on the figure as the coloured red circle on the red curve). The False Positive Rate and True Positive Rate is also presented for  the \citealp{whitaker_large_2012} $UVJ$ selection (blue star). \textit{Middle column}: Same as left column but at $z=3$. \textit{Right column}: Same as left column but at $z=4$.}
    \label{fig:shark_fig}
\end{figure*}

\noindent To calculate the $p(q)$ threshold above which a galaxy is classed as quiescent and assess the performance of the method, we use galaxies in the {\sc Shark} simulation (Section \ref{sec:shark}). At redshift snapshots of $z=2, z=3$, and $z=4$ we select a massive galaxy sample (log(M$_{*}/$M$_{\odot})>10$). We then define QGs in the sample as those with specific star formation rates log$({\rm sSFR}$ t$_{H} \mathrm{yr}^{-1}) < 0.2$, where $t_{H}$ is the age of the universe at that redshift snapshot and the sSFR is averaged over the last $\sim50$ Myr. We fit the GMM to the $NUV-U$, $V-J$ and $U-V$ colours at each redshift snapshot. We then calculate the Receiver Operator Characteristic (ROC) curve, which is simply the False Positive Rate (FPR) as a function of True Positive Rate (TPR). The Area-Under-Curve (AUC) can be used as a measure of effectiveness of the classification method: an AUC very close to unity shows that the method correctly classifies objects with a high TPR and low FPR. We also compute the same numbers for the \citealp{whitaker_large_2012} $UVJ$ selection method (defined as $U - V > 0.8 \times (V - J) + 0.7$, $U - V > 1.3$ and $V - J < 1.5$). Figure \ref{fig:shark_fig} shows both the $NUVU$-$VJ$ and $UVJ$ diagrams and the ROC curves for each redshift snapshot. The AUC increases from $z=2$ to $z=4$ , meaning that the GMM has increasing chance of correctly identifying QGs ($88\%$, $93\%$ and $97\%$ at z$=2,3,4$ respectively). It is not possible to calculate the AUC for $UVJ$ because there is only one selection ``threshold", which is whether or not a galaxy falls inside the $UVJ$ quiescent area. We can instead calculate the TPR and FPR for $UVJ$. The optimal $p(q)$ threshold is defined as the maximum of the geometric mean, which finds the threshold at which the TPR and FPR are perfectly balanced (or the top left most point on the ROC curve), and is defined as $max(\sqrt{TPR*(1-FPR)}$. The resulting $p(q)$ thresholds at $z=2$, $z=3$, and $z=4$ are calculated using the geometric mean and are shown in Table \ref{tab:pq_table}. 

\subsubsection{UVJ versus GMM in \textsc{SHARK}}

\noindent At $z=2$, $UVJ$ performs similarly to $NUV-VJ$ at $p(q)\gtrapprox0.1$, with $UVJ$ having FPR=$30\%$, TPR=$90\%$ and $NUV-VJ$ with FPR=$33\%$, TPR=$89\%$. However, the optimal threshold for $NUV-VJ$ performs substantially better at this redshift, with a TPR only a few $\%$ lower ($81\%$), whilst the FPR is more than halved, reducing from $33\%$ to $15\%$. If we consider instead the contamination, which is defined as the number of galaxies defined as quiescent compared to the total number of UVJ quiescent galaxies, the conclusion changes. The contamination fraction of UVJ in \textsc{SHARK} at $z=2,3,4$ is $80,86,91\%$, further highlighting the issues with UVJ. Looking to higher redshift, it is evident that $NUV-VJ$/ GMM classification not only out-performs $UVJ$ classification, but also substantially increases in effectiveness with increasing redshift, highlighting the usefulness of the method especially at $z>3$.   

\begin{table}
    \caption{Quiescent probability ($p(q)_{50}$) thresholds for redshift ranges, defined by the maximum of the geometric mean of the ROC curve trained on {\sc Shark} simulated data.}
    \centering
    \begin{tabular}[t]{cc}
    \toprule
        Redshift &  ${p(q)_{50}>}$  \\
    \midrule 
       $2.0<z<3.0$  & $0.56$\\
        $3.0<z<4.0$  & $0.18$\\
         $4.0<z<5.0$  & $0.01$\\ 
        \bottomrule
    \end{tabular}
    \label{tab:pq_table}
\end{table}

\section{Application to COSMOS2020}\label{results}
\noindent We apply the GMM to the massive galaxy sample in COSMOS2020 at $2<z<5$ selected in Section \ref{qsel}. Whilst \textsc{SHARK} produces galaxies at redshift snapshots of $z=2,3,4$, observed galaxies are continuously distributed in redshift.
We therefore choose not to assume that the $p(q)$ threshold is a step function, 
but rather a smoothly varying function of redshift. We fit a second order polynomial to the three points $(z{phot},p(q))=(2,0.56;3,0.18;4,0.01)$ and use this function to determine if a galaxy is chosen as quiescent or not. This naturally incorporates the assumption that as we move to higher redshift, in order to be defined as quiescent, galaxies may look less like quiescent galaxies at $z=2$, and more like post-starburst galaxies. We require that at $z=2-4$, the median of the $p(q)$ distribution is above the threshold. At $z\geq4$, we instead require that $95\%$ of the $p(q)$ distribution is above $p(q)=0.01$, in order to only select the highest confidence candidates. The application of redshift dependent $p(q)$ thresholds selects 1455 QGs at $2<z<3$ and 230 QGs at $3<z<5$. We visually inspect cut-outs of the $3<z<5$ sample in the optical and NIR bands, and remove nine QG candidates which are spurious, reducing the total $3<z<5$ sample to 221 in number. In the following sections, we present the properties of our sample of QGs, compare its statistics with those of galaxies selected using traditional colour diagrams, and discuss the differences between methods. 

\subsection{Comparing GMM/$NUV-VJ$ to $UVJ$ and NUVrJ colour selections} \label{comparing}

\begin{figure*}
    \centering
    \includegraphics[width=0.9\textwidth]{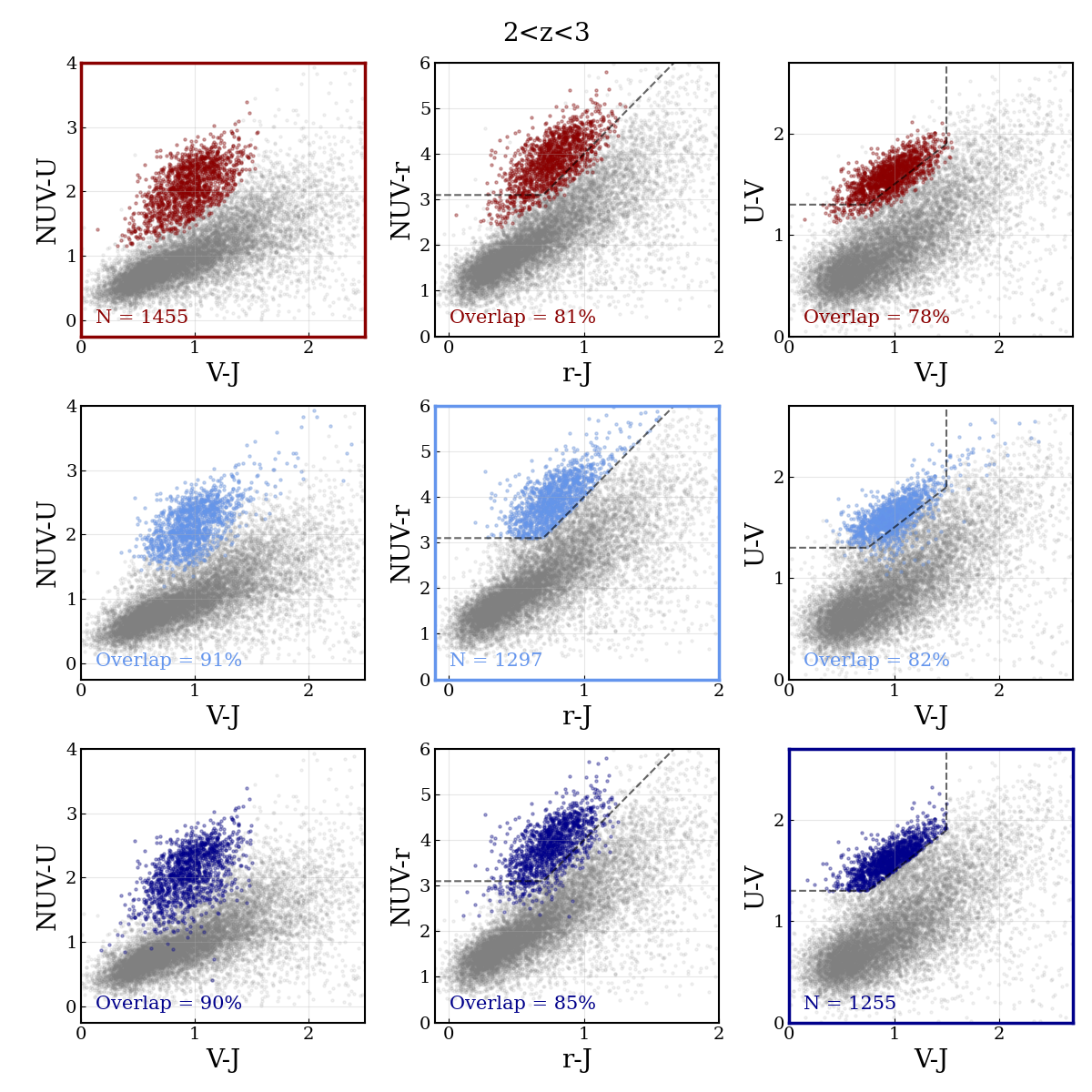}
    \caption{\textit{Top row}: $NUV-VJ$ selected QGs at $2<z<3$ (dark red, left figure) shown in $NUV-VJ$ space, $UVJ$ space and NUVrJ space. \textit{Middle row}: NUVrJ selected QGs (middle figure, light blue) shown in $NUV-VJ$ and $UVJ$ colour diagrams. \textit{Bottom row}: $UVJ$ selected QGs (right figure, dark blue) shown in $NUV-VJ$ and NUVrJ colour diagrams. For each panel, the overlap is given, defined as the number of galaxies selected by that selection method divided by the number of galaxies selected by the main selection method for that row. In the first row we show the GMM selected sample (dark red) at $2<z<3$ in the $NUV-VJ$ diagram, NUVrJ diagram and $UVJ$ diagram.  The second row shows the NUVrJ selected sample in NUVrJ (light blue) as well as the other two colour colour diagrams and the third row shows the $UVJ$ selected sample in $UVJ$ as well as in the other two colour colour diagrams (dark blue). }
    \label{fig:colour1}
\end{figure*}

\begin{figure*}
    \centering
    \includegraphics[width=0.9\textwidth]{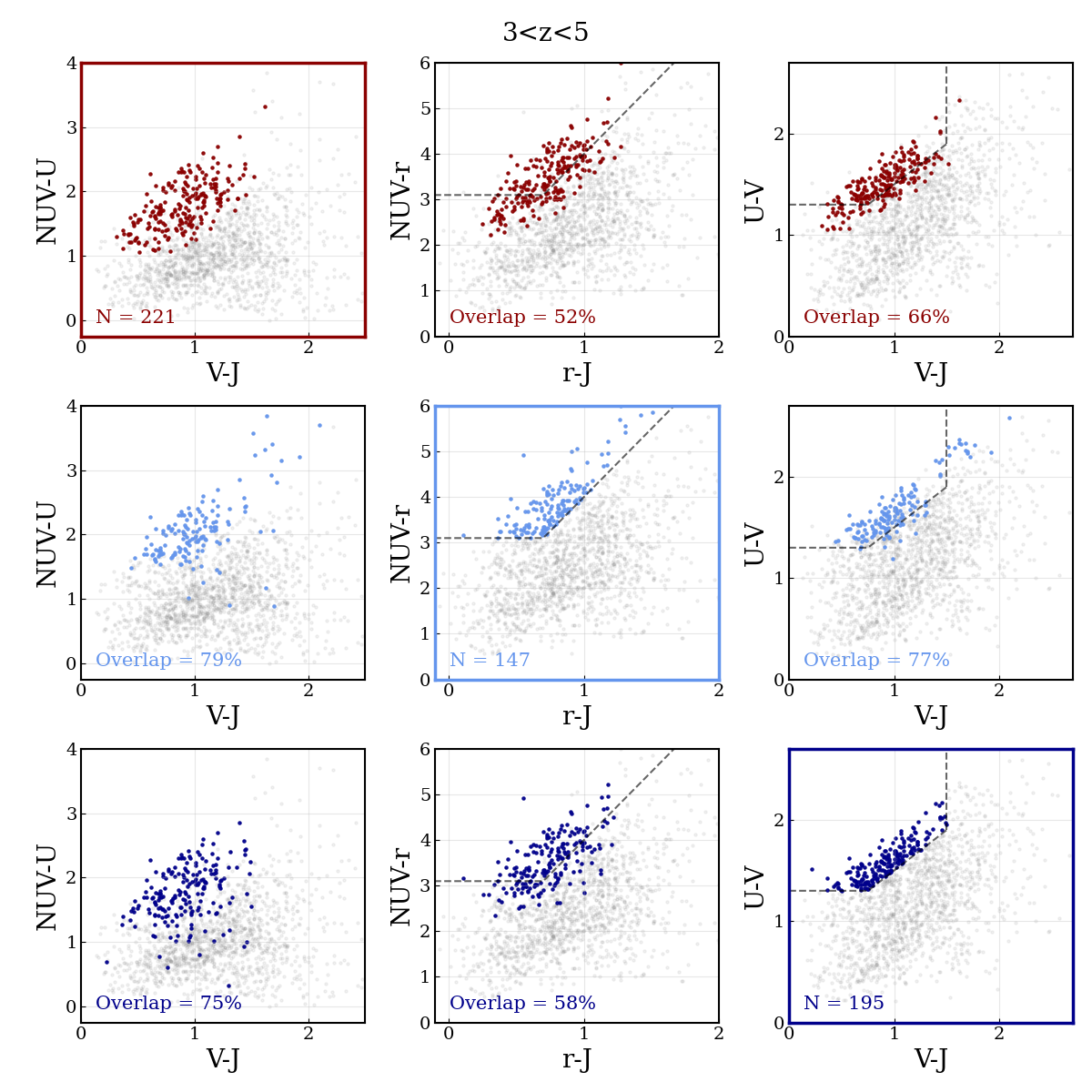}
    \caption{Same as figure \ref{fig:colour1} but for $3<z<5$.}
    \label{fig:colour2}
\end{figure*}

\noindent We apply both the \cite{whitaker_large_2012} $UVJ$ selection as well as the \cite{ilbert_mass_2013} $NUVrJ$ selection for quiescence to our massive galaxy samples at $2<z<3$ and $3<z<5$, and compare these traditional colour selections to the GMM colour selection method. For each panel of Figure \ref{fig:colour1}, we calculate the overlap between selections, which is defined as the sample size divided by the sample size of the ``main'' selection for that row, which is highlighted by the coloured panel. 

At $2<z<3$, the overlap between all three selection methods is high: $UVJ$ and $NUVrJ$ selections have an $\sim85\%$ overlap, whilst the GMM selected sample also agrees well, sharing $\sim80-90\%$ with $UVJ$ and $NUVrJ$ selected QGs. This would be expected if the rest frame fluxes were measured directly from the template, but in this case they are weighted by the observed fluxes and so the conclusion is that the different selections are generally measuring the same \textit{observed} SED shape. \cite{hwang_revisiting_2021}, who compared $NUVrJ$ and $UVJ$ selected samples at $0<z<3$ in COSMOS2015, found a similar overlap ($\sim85\%$). This suggests that the choice of rest-frame colour diagram selection at $z<3$ is not crucial, although modifications to both $UVJ$ and $NUVrJ$ colour selections may be in order.\\ 

At $3<z<5$ the agreement between colour selections is less clear: $UVJ$ and $NUVrJ$ share $58-77\%$ of the same sample, whilst GMM shares $52-79\%$ with $NUVrJ$ and $66-75\%$ with $UVJ$. This difference is highlighted clearly in Figure \ref{fig:colour2}. This is likely due to two reasons: firstly, that neither $UVJ$ nor $NUVrJ$ selections include the region where post-starburst galaxies are most often found, which is at the lower edge beyond the $U-V<1.3$ boundary. Secondly, the contamination fraction of $UVJ$ and $NUVrJ$ is likely much higher than the GMM selection method. Whilst it is easy to remedy the first issue in both cases by simply lowering or removing the dividing line in $U-V/NUV-r$ as discussed above, this may also increase the contamination of the sample by introducing $U-V/NUV-r$ blue star forming galaxies. This is something that \cite{schreiber_near_2018} confirmed in their analysis. For completeness, we report the performance of an extended $UVJ$ selection bench-marked against SHARK in Appendix \ref{app:extended_uvj}. The issue of contamination is more difficult to solve due to the nature of these colour selections, however our method aims to alleviate this problem by both the introduction of the third $NUV-U$ colour, and the option of choosing galaxies based on a probability distribution ideally makes it easier to sharpen the boundaries between different populations.

\subsection{Full spectral energy distribution}

\begin{figure*}
  \centering
  \begin{minipage}[b]{1.0\textwidth}
    \includegraphics[width=\textwidth]{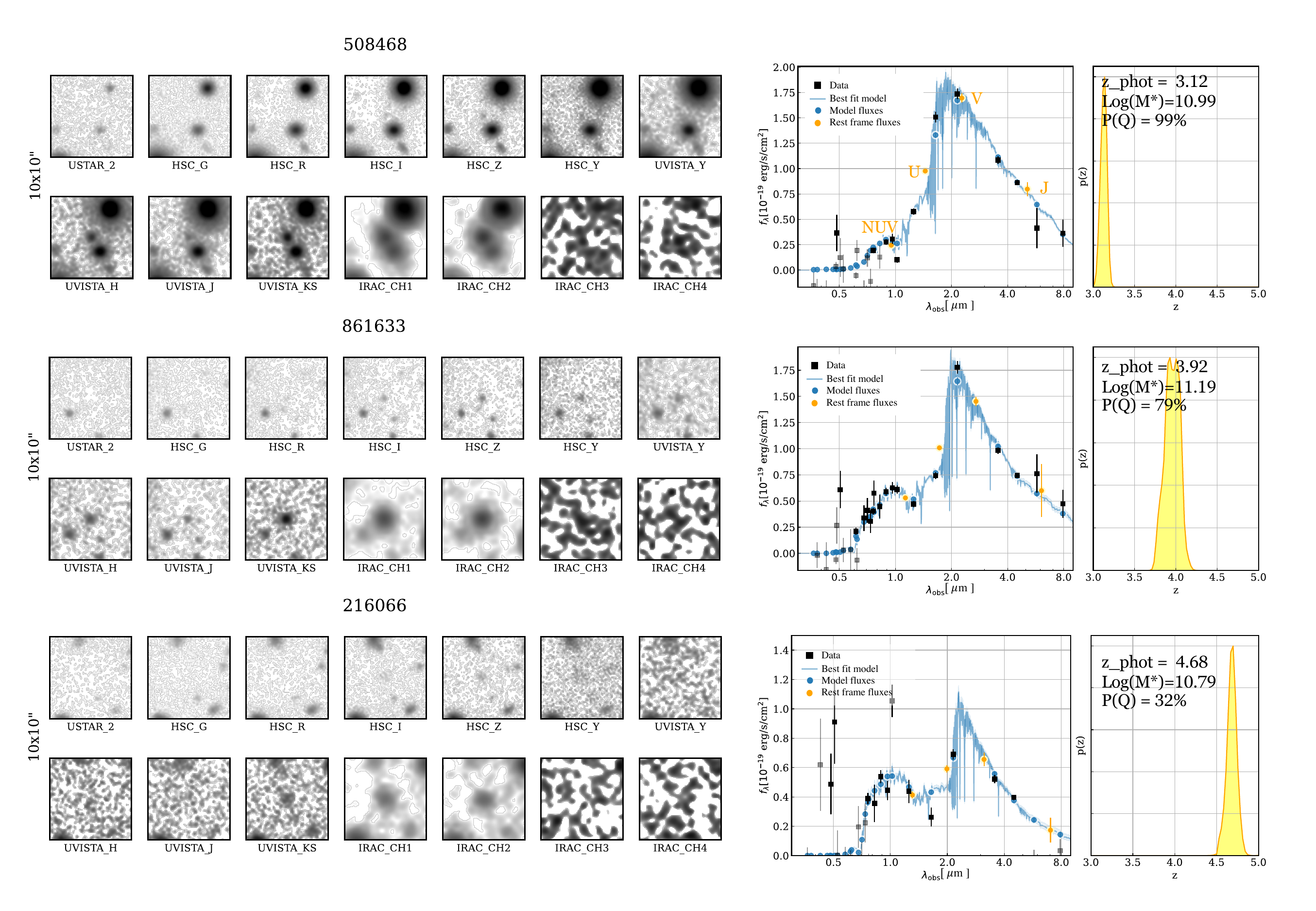}
  \end{minipage}
  \caption{$10\times10"$ Postage stamp cutouts from selected bands in the optical to NIR scaled by $\pm3$RMS (\textit{left}) and corresponding best fit SEDs (\textit{right}) for three quiescent galaxy candidates selected using the $NUV-VJ$ method, taken from the main sample. The observed photometric points are shown as black circles (any with SNR$<2$ in grey) and the best fit model is shown in blue. The rest frame $NUV,U,V,J$ photometry is shown by the orange circles. The $p(z)$ is shown to the right of each SED in yellow, along with its best-fit photometric redshift, stellar mass, and quiescent probability median value ($P(q)_{50}$).}
  \label{fig:egseds}
 \end{figure*}

\begin{figure*}[!tbp]
  \centering
  \begin{minipage}[b]{0.90\textwidth}
    \includegraphics[width=\textwidth]{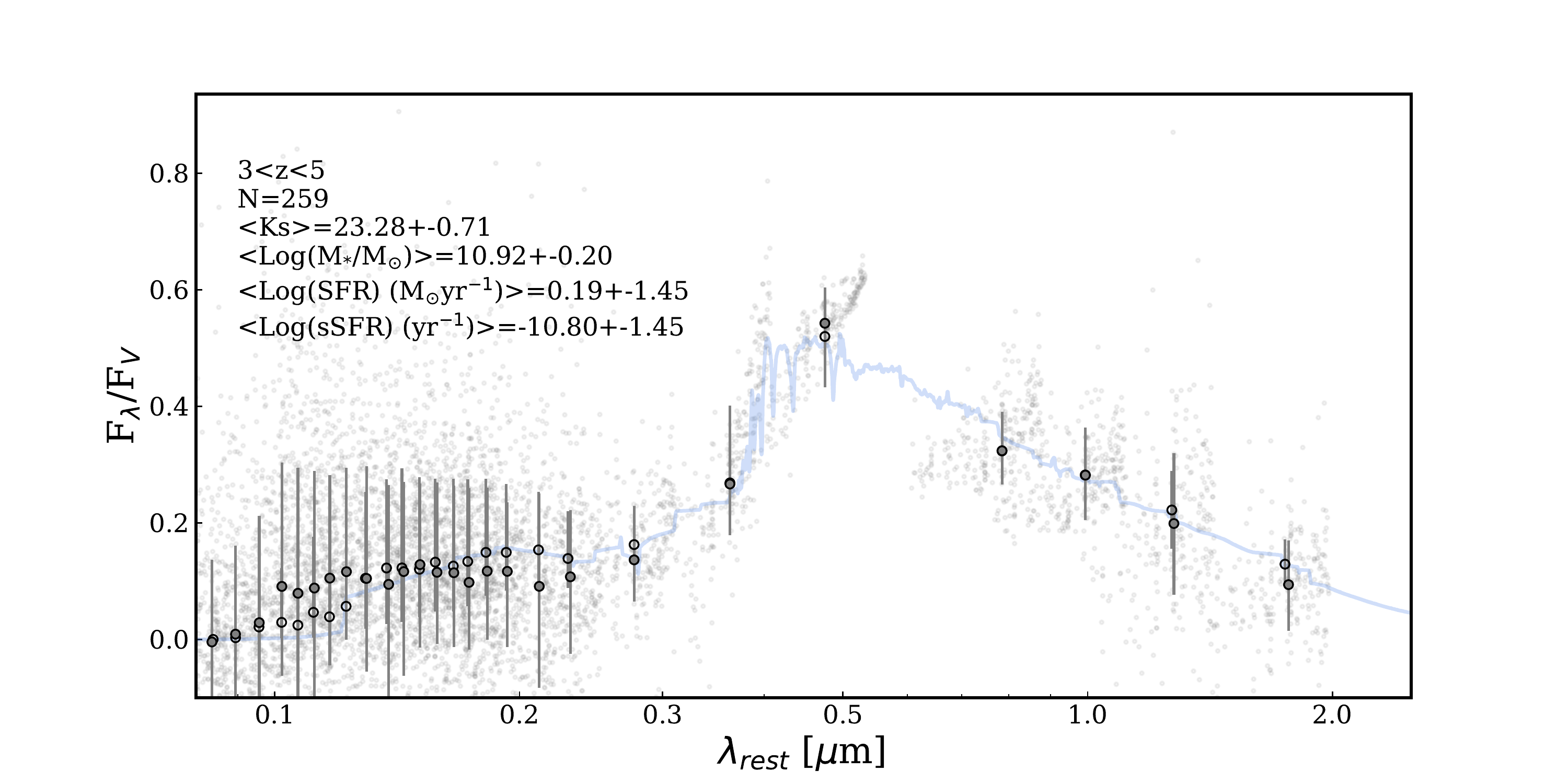}
  \end{minipage}

  \caption{\textit{Top}: Median smoothed rest frame SED for the GMM/$NUV-VJ$ selected quiescent sample at $3<z<5$. Rest frame photometry for all galaxies is shown in grey, whilst the median per band and the associated $16$th and $84$th percentiles (error bars) are shown in dark grey. The median best fit (smoothed) model is shown in red. The median model photometry per band is shown by the open face circles. The location of the sample in $NUV-VJ$ space is shown in the top right inset, whilst at the top left the sample median properties (number, median $K_{s}$ band magnitude, median log(stellar mass), median log(SFR), median log(sSSFR)) are shown, as well as the standard deviation.}
  \label{rf_median_sed}
\end{figure*}

\noindent To offer a more complete view of the full SEDs of our selected QGs, in Figure \ref{fig:egseds} we show both the cutouts from optical/NIR bands and best-fit SEDs for three candidates at $z_{phot}=3.12$, $3.92$, and $4.68$. We also compute the median rest-frame photometry and best-fit models for the whole GMM-selected sample normalized to $V=1$ (Figure \ref{rf_median_sed}). 
The median values of $K_{s}$ observed magnitudes, \mstar, SFRs, and sSFRs are also reported in the figure.\\ 

The galaxies in our sample show a strong Balmer/$4000$ \AA\ break indicating a dominant aged stellar population on average, as expected, 
However, they still have residual flux blue-ward of the break, indicative of a younger populations and recent quenching.
The median mass (log(M$_{*}/$M$_{\odot})=10.92$) of this sample is $\sim0.3$ dex higher than the mass cut for the original sample selection, confirming that only the most massive galaxies have quenched their star formation at $z>3$. The median log(sSFR$/\mathrm{yr}^{-1}$)$=-10.8$ indicates that these galaxies are best fit mainly by templates that include little or no ongoing star formation. 

\subsection{Spectroscopically confirmed QGs}
\label{sec:spectroscopy}
\noindent As a confidence check, we cross-matched our sample of candidate QGs with a literature compilation of 7 spectroscopically confirmed $z\gtrsim3$ QGs in COSMOS from \citealp{forrest_massive_2020} (4), \citealp{forrest_massive_2020}/\citealp{marsan_spectroscopic_2015}/\citealp{saracco_rapid_2020} (1), \citealp{valentino_quiescent_2020} (1) and \citealp{deugenio_hst_2020} (1). For 6/7 sources we retrieve $p(q)\gtrsim10$\%, consistent with being QG according to our selection. To the remaining galaxy at $z_{spec}=3.352$ \citep{marsan_spectroscopic_2015, forrest_massive_2020, saracco_rapid_2020}, we assign $p(q)=0.6$\%. This source would thus not be selected using our fiducial threshold at $3<z<4$. We note that this galaxy has experienced rapid quenching, possibly due to an AGN. The presence of the latter is inferred from the large [OIII]/H$\beta$ emission line ratio, with the oxygen line possibly contaminating the $K_{s}$ photometry (see the discussion in \citealt{forrest_extremely_2020}).

\section{Number densities} \label{ndensities}

\begin{figure*}
    \centering
    \includegraphics[width=0.9\textwidth]{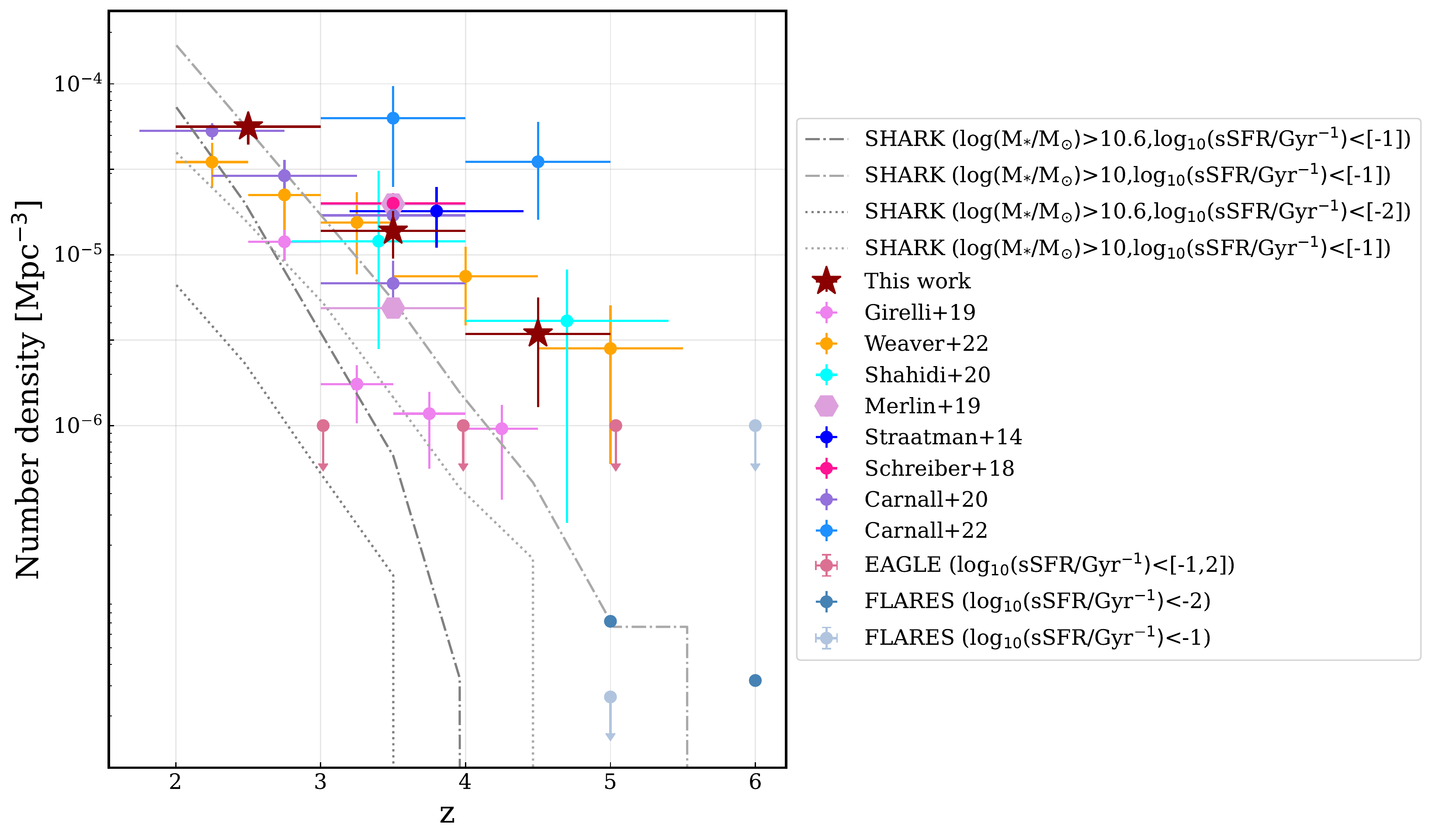}
    \caption{Number densities of massive QGs as a function of redshift at $2<z<5$. We report the number densities calculated for our sample (dark red stars) in redshifts bins of $2<z<3$ (log(M$_{*}/$M$_{\odot})>10$) , $3<z<4$ (log(M$_{*}/$M$_{\odot})>10.6$) and $4<z<5$ (log(M$_{*}/$M$_{\odot})>10.6$). The errors on the number densities include both Poisson noise and cosmic variance added in quadrature. We plot number densities from \citealp{weaver_cosmos2020_2022}, \citealp{shahidi_selection_2020}, \citealp{straatman_substantial_2014}, \citealp{schreiber_near_2018}, \citealp{carnall_timing_2020},\citealp{merlin_red_2019}, the extended sample from \citealp{girelli_massive_2019} and \citealp{carnall_first_2022}. Also shown are number densities for the \textsc{FLARES} and \textsc{EAGLE} simulations for massive galaxies (log(M$_{*}/$M$_{\odot})>10.6$) at two quiescent selection thresholds (log$_{10}$(sSFR/Gyr$^{-1})<$[-1,-2]) \citep{lovell_flares_2022}, as well as quiescent galaxies from the \textsc{SHARK} simulation at two mass (log(M$_{*}/$M$_{\odot})>[10.,10.6]$) and sSFR selection thresholds (log$_{10}$(sSFR/Gyr$^{-1})<$[-1,-2])}
    \label{fig:numberdens}
\end{figure*}

\noindent The number densities of massive QGs at $z>3$ is an important constraint on galaxy evolution and theory. The assembly of galaxies with such high stellar masses in an evolved state within only 1.5-2 billion years not only places important constraints on the formation of the first galaxies (e.g. \citealp{steinhardt_impossibly_2016} and references within), but also on the cosmic star formation rate density (cSFRD) \citep{merlin_red_2019}. As such, the number density of these galaxies provides important context to early galaxy evolution. However, the number densities of massive QGs at $z>3$ has been an intense topic of debate, due to both the disagreement between observations and theory, and between the observational studies themselves.\\ 

Number densities derived from photometric and spectroscopic observations span $\sim2$ dex, but they generally remain higher than those extracted from simulations \citep[Valentino et al., in press]{ilbert_mass_2013, straatman_substantial_2014, davidzon_cosmos2015_2017, schreiber_near_2018, merlin_chasing_2018, merlin_red_2019, girelli_massive_2019, shahidi_selection_2020, carnall_timing_2020, santini_emergence_2020, valentino_quiescent_2020, carnall_first_2022}. The disagreement among observational works and with theoretical predictions is likely due to a mix of several factors, primarily different sample selections and quiescent criteria used together with analyses done using multiple diverse data sets. Of particular note is the size of such fields, as statistics can be affected by both the rarity of these galaxies and by cosmic variance.

The latter is strongly mitigated by the large contiguous area of the COSMOS field. Here we compute and present the number densities for our sample of QGs 

in three bins at $2<z<3$, $3<z<4$ and $4<z<5$ (Table \ref{dense_table}). Number densities are calculated using the survey area of the combined \textit{HSC}/\textit{UVISTA} coverage, which corresponds to 1.27 square degrees. We calculate the fractional error due to cosmic variance using the method of \cite{steinhardt_finding_2021} who extend the method of \cite{moster_cosmic_2011} for use in the early universe (for details, see \citealp{steinhardt_finding_2021} and \citealp{weaver_cosmos2020_2022}). This is combined in quadrature with the Poisson error to get the total error budget.\\

In Figure \ref{fig:numberdens}, we show our number densities along with a compilation of number densities from other observational studies (e.g., Valentino et al., in press and references therein).
Additionally, we show QG number densities from the \textsc{SHARK} simulation at two mass (log(M$_{*}/$M$_{\odot})>[10.,10.6]$) and sSFR selection thresholds (log$_{10}$(sSFR/Gyr$^{-1})<$[-1,-2]), as well as both the \textsc{flares} ($z\geq 5$) and \textsc{eagle} ($z\leq 5$) simulations for QGs at (log(M$_{*}/$M$_{\odot})>10.6$) at two different selections (log$_{10}$(sSFR/Gyr$^{-1})<$[-1,-2]) \citep{lovell_flares_2022}. It is important to note that the literature compilation of QGs at $z>3$ comprise selections at different mass limits ranging from (log(M$_{*}/$M$_{\odot})>10$, \citep{carnall_timing_2020} upwards, whilst our mass limit is much higher (log(M$_{*}/$M$_{\odot})>10.6$). 
In general, our number densities at $3<z<4$ agree with the observational studies within $1\sigma$ errors. It is interesting to note that the number densities derived by \cite{girelli_massive_2019} are much lower than those derived from COSMOS2020, which likely arises from the COSMOS2020 catalog including all UVISTA stripes, some of which have known over-densities (e.g. \citealp{mcconachie_spectroscopic_2021}). Looking to higher redshift at $z>4$, our estimates agree with those of \cite{weaver_cosmos2020_2022} and \cite{shahidi_selection_2020}. Generally, it appears that number densities for massive QGs derived from a variety of different fields, selected with a variety of methods, are finally converging on agreement.\\

 Previously, simulations were not able to reproduce the number densities of massive QGs at $z>3$ observed in the real universe by a factor of 1-2 dex; this tension remains slightly with \textsc{eagle}, which  predicts only upper limits for QGs at log(M$_{*}/$M$_{\odot})>10.6$ that are several times lower than most observations.
 At $z=5$, the results from \textsc{flares} are $\sim50\times$ smaller than our current limits at fixed \mstar\ threshold, resulting in a direct tension. At $z>5$, this simulation predicts similarly low number densities, but the lack of data from observations at $z\gtrapprox5$ means that for the first time at high redshift, simulations have QGs where observations have found none. \textsc{SHARK} performs the best at producing similar number densities of massive QGs to observations up to at least $z\sim3$, but has a dearth of massive (log(M$_{*}/$M$_{\odot})>10.6$) QGs at $z\gtrsim4$. The only QGs at $z>4$ in \textsc{SHARK} are $\sim4$ times less massive than those in observations, and in fewer numbers too, implying that those galaxies which have been able to quench are still to accumulate mass. Finding QGs at $z\gtrapprox5$ will require both the combination of wide optical/NIR surveys to find such rare galaxies, and also deep NIR/MIR spectroscopy to confirm them.

\begin{table}%[ht]
    \centering
    \begin{tabular}[t]{lc}
    \toprule
        Redshift &  $N$ Mpc$^{-3}$\\
    \midrule 
        $2.0<z<3.0$ & $5.62\pm1.2$x$10^{-5}$ \\
       $3.0<z<4.0$ & $1.38\pm0.4$x$10^{-5}$\\ 
        $4.0<z<5.0$ & $3.45\pm2.16$x$10^{-6}$ \\ 
        \bottomrule
    \end{tabular}
    \label{dense_table}
    \caption{Number densities for galaxies in our sample at $2<z<3$ log(M$_{*}/$M$_{\odot})>10$, $3<z<4$ log(M$_{*}/$M$_{\odot})>10.6$ and $4<z<5$ log(M$_{*}/$M$_{\odot})>10.6$. Errors are calculated by combining in quadrature the Poisson uncertainty and the fractional error due to cosmic variance.}
    \label{tab:ndens_table}
\end{table}

\section{What fraction of QGs at $3<z<5$ are post-starburst?}\label{psb-section}

\noindent In the epoch of the universe where quenching begins, it may make more sense to instead differentiate between quenching, recently quenched, and older (or ``true'') QGs. This issue confronts a more philosophical question in extra-galactic astrophysics: what is the definition of a quiescent galaxy? There appear to be two ways to deal with this population exhibiting a wide variety of quenching states: firstly, one can split the entire sample into young quiescent/PSB/recently quenched, and old quiescent (e.g. \citealp{ichikawa_recently_2017, schreiber_near_2018, park_rapid_2022}). Alternatively, the entire sample can be treated as homogeneous, which is what we have done with this analysis. The prevalence of QGs with recent star formation at $z>3$ - the so-called PSB population - has been measured by a plethora of works, both from photometric and spectroscopic data. Whilst at low redshift, the definitions of post-starburst, albeit broad, are based on measurements involving spectral indices or spectra-based PCA that encode information about strong Balmer absorption lines as well as weak or no emission lines (e.g. \citealp{wild_new_2014}, \citealp{chen_post-starburst_2019}, \citealp{wild_star_2020}, \citealp{wilkinson_starburst_2021}), this type of definition is not possible to apply to large photometrically selected samples. We instead consider how one might define a PSB galaxy in a sample of photometrically selected QGs based just on colours.\\ 

\cite{marsan_number_2020} proposed a definition of PSB based on a visual SED inspection and the presence of a UV peak brighter than the emission red-ward of the Balmer/4000\AA break. This study estimates that PSB galaxies comprise $28\%$ of the massive (log(M$_{*}/$M$_{\odot})>11$) population at $3<z<4$ and $17\%$ at $4<z<6$. Approximately twice higher fractions of PSB at log(M$_{*}/$M$_{\odot})>11$ are reported by \cite{deugenio_hst_2020} based on their sample of spectroscopically confirmed objects at $z>3$ with measured $D_{n4000}, D_{B}$ and H$_{\delta A}$ indices.

Thirty percent of the spectroscopic sample in \cite{forrest_massive_2020} ceased star formation less than 300 Myr prior the epoch of the observed redshift at $3<z<$. At the same epoch, \cite{schreiber_near_2018} report that ``young quiescent'' galaxies, defined as $UVJ$-star forming galaxies with $V-J<2.6$ but below some sSFR threshold, are just as numerous as classical QGs.\\

Here we base a definition of PSB on the strength of the ``blue bump'' in the SEDs traced by the $NUV-U$ colour.
The distribution of $NUV-U$ colours for our QG sample of 221 galaxies peaks at $NUV-U=1.75$. Only $28$\% of galaxies have $NUV-U>2$ and only one galaxy has $NUV-U>3$. Based on our single stellar population model from Section \ref{newcolor}, a stellar population will not reach $NUV-U>2$ until $0.5-1$ Gyr has passed and $NUV-U>3$ until well after 1 Gyr, assuming no dust attenuation. Based on this, we can define PSBs as those galaxies in our sample with $NUV-U<2$, comprising $70$\% of QGs at $3<z<4$ and $87$\% at $4<z<5$. This is marginally higher than what previously reported \citep{schreiber_near_2018}. 
If we restrict our selection to only the most massive galaxies, i.e., those with log(M$_{*}/$M$_{\odot})>11$, then these numbers change to $60$\% at $3<z<4$ and $86$\% at $4<z<5$. This agrees with the $60-70$\% value given by \cite{deugenio_hst_2020} for galaxies at $z\sim3$. As a fraction of the entire ultra-massive galaxy population, PSBs only comprise $7$\% at $3<z<4$, and $12$\% at $4<z<5$. Together with the results of the previous section, it is clear that $3<z<5$ is an active era in the history of the universe, one in which 
the transition to quiescence is fairly common.\\ 

\section{Will we find old quiescent galaxies at $z>4$ with $JWST$?}

QGs have already been spectroscopically confirmed with $JWST$ \citep{nanayakkara_population_2022, carnall_massive_2023}, however no old quiescent galaxies have been found at $z>4$ yet. In general, the blue colours of our $3<z<5$ QG candidates indicates a lack of very red, evolved galaxies such as the one presented in \citealp{glazebrook_massive_2017}. This could be because as an ensemble, the QG population at $z>3$ has only recently quenched, and older QGs are much rarer. However, at $z>4.5$, the reduced sensitivity to older/redder quiescent galaxies due to the lack of data directly covering the $4000\AA$ break could be biasing us to the bluest (and youngest) quiescent galaxies. Although this is easily side-stepped with $JWST/NIRSPEC$, it is difficult to solve even with $JWST$ broad-band $NIRCAM$ imaging; at $4.5<z<5$, the $4000\AA$ break falls at $2.2-2.4\mu m$ and is straddled by $F200W$ on the blue side and $F277W$ (or redder bands such as $F350W$) on the red side (depending on the program), with no observations of quiescent galaxies at $z>4.5$ currently spanning the midpoint or upper edge of the break (see e.g. Valentino et al., in press, and \citet{carnall_first_2022}). This could easily be alleviated with the inclusion of $NIRCAM$ narrow-bands covering 2.2-2.4$\mu m$, such as $F210M$ or $F250M$. Currently, the only Cycle 1 program that has the combination of $F200W$, $F277W/F350W$ and at least one medium band filter in between is the Canadian NIRISS Unbiased Cluster Survey (CANUCS; PI Willot \citep{willott_canucs_2017}), which has both $F210M$ and $F250M$. However, the combined area covered by these filters is $\sim50$ arc-minutes squared, which would net less than one massive quiescent galaxy based on our number densities (see Section \ref{ndensities}), assuming the same space distribution on the sky. In short, one would have to be incredibly lucky to find an evolved quiescent galaxy at $z>4.5$ without both area and either spectroscopy or medium-band photometry.

\section{Summary and conclusions}\label{con}

\noindent In this work, we explore the massive quiescent galaxy population at $3<z<5$ in $COSMOS$ using the latest photometric catalogue, $COSMOS2020$. We create a new catalogue using the $COSMOS2020$ photometry, with \textit{Spitzer/IRAC} errors boosted and derive redshifts and physical parameters with \ezp.\ We motivate the need for a new rest-frame color diagram to select for QGs at $z>3$ and present the Gaussian Mixture Model (GMM)/$NUVU-VJ$ method as a viable alternative. We use a GMM to fit the $NUV-U,U-V,V-J$ colours of massive galaxies at $2<z<3$ and galaxies are assigned a probability of being quiescent based on their bootstrapped rest frame colours. This GMM model is then applied to the colours of massive galaxies at $3<z<5$. Both the GMM model and code to calculate quiescent probabilities from rest frame flux densities are made available online \footnote{\url{https://github.com/kmlgould/GMM-quiescent}}. Our key findings can be summarised below: 

\begin{itemize}
    \item We calculate the quiescent probability for simulated galaxies from the \textsc{shark} semi-analytical model at redshift snapshots of $z=2, 3$, and $4$ and find that GMM performs just as well as classical $UVJ$ selection at $z=2$, with both methods returning similar true positive and false positive rates (TPR$\sim90\%$, FPR$\sim30\%$). However, at $z>3$, the GMM method outperforms classical $UVJ$ selection and particularly excels at $z\geq4$, where the TPR is almost $100\%$ compared to $UVJ$ that has less than $40\%$ TPR. This highlights our proposed method as a viable alternative to traditional colour selection methods at $z>3$.
    
    \item We compare our GMM-selected quiescent galaxies (QGs) in $COSMOS2020$ to traditional rest-frame colour selected quiescent galaxies ($UVJ$, $NUVrJ$) and calculate the overlap between selection methods. We find that all three colour selections have high agreement at $2<z<3$, with methods selecting $\sim85\%$ of the same galaxies, implying that the exact choice of rest-frame colour selection method at $2<z<3$ is not crucial. However, overlap between colour selections at $z>3$ reduces to $\sim65\%$, implying that one should be careful about using $UVJ$ and $NUVrJ$ colour selections interchangeably. Generally, the bluer colours of GMM selected QGs at $3<z<5$ implies a dearth of any highly evolved QGs in this epoch.

    \item We compute the number densities of QGs at $2<z<5$ and confirm the overall abundance of QGs presented in many different works, which our values agree with within the uncertainties. Given the conservative nature of this work (the boosting of $IRAC$ errors, the high mass limit and the probabilistic approach), we interpret the number densities as a lower limit. 

\end{itemize}

\noindent The next few years will undoubtedly see our knowledge of QGs at $z>3$ grow vastly with the recent launch and successful commissioning of \textit{JWST} \citep{rigby_characterization_2022}, for which QGs at $z>3$ have already been reported \citep{carnall_first_2022, perez-gonzalez_ceers_2022,cheng_jwsts_2022, rodighiero_jwst_2023} Valentino et al., in press and even confirmed \citep{nanayakkara_population_2022, carnall_massive_2023}, as well as ongoing large ground-based surveys such as the Cosmic Dawn Survey (\citealp{moneti_euclid_2022}, McPartland et al., 2023, in prep.). Spectroscopic confirmation of multiple QG candidates with, e.g., \textit{NIRSPEC} can be done in a matter of minutes \citep{glazebrook_how_2021, nanayakkara_population_2022}, and whilst $\sim8$ hours with $VLT/X-shooter$ or $Keck/MOSFIRE$ are only sufficient to obtain a redshift and stellar populations, the same time (or less) with \textit{JWST} could provide the required S/N to study detailed physics such as velocity dispersions and metallicities \citep{nanayakkara_massive_2021, carnall_massive_2023}. Whilst \textit{JWST} will be crucial for confirming quiescent galaxy candidates and studying their physical properties in detail, wide area photometric surveys still have an important role to play, both for the selection of, and statistical study of these galaxies.  

\begin{acknowledgments}
We acknowledge the constructive comments from the referee, which significantly improved the content and presentation of the results. The Cosmic Dawn Center (DAWN) is funded by the Danish National Research Foundation under grant No. 140. S.J. acknowledges the financial support from the European Union's Horizon research and innovation program under the Marie Sk\l{}odowska-Curie grant agreement No. 101060888.
\end{acknowledgments}

\vspace{5mm}

\software{eazy-py \citep{brammer_eazy-py_2021}, scipy \citep{virtanen_scipy_2020}, numpy \citep{harris_array_2020}, matplotlib \citep{hunter_matplotlib_2007}, sci-kit learn \citep{pedregosa_scikit-learn_2011}, astropy \citep{collaboration_astropy_2013,astropy_collaboration_astropy_2018}, python-fsps \citep{conroy_propagation_2010, conroy_propagation_2009}.}

\appendix

\section{Rest-frame flux densities}\label{rf-fluxes}

Consider two methods for inferring the flux density of a given galaxy at redshift, $z$, in a rest-frame bandpass, e.g., the $V$ (visual) band in the Johnson filter system \citep{johnson_photometric_1955,bessell_ubvri_1990},  with a central wavelength\footnote{The pivot wavelength is a convenient source-independent definition of the central wavelength of a bandpass \citep{tokunaga_mauna_2005}} $\lambda_c\sim5500~\mathrm{\AA}$.  In the first method we identify two observed bandpasses with central wavelengths that bracket the redshifted rest-frame bandpass $\lambda_{c,A} < \lambda_c (1+z) < \lambda_{c,B} $ and perform a linear interpolation between the flux densities in the observed bandpasses $A$ and $B$.  In a second method we simply adopt the $V$ band flux density of the best-fit model template that was used to estimate the photometric redshift of the galaxy.  The first method has the benefit of being purely empirical, though it will be limited by noise in the two photometric measurements used for the interpolation.  The second method incorporates constraints from all available measurements, but it will be more sensitive to systematic biases induced by the adopted model template library.

Here we adopt a methodology that is a hybrid between these two extremes that first makes better use of multiple observational constraints and second somewhat relaxes the dependence on the template library.  Specifically, for a galaxy at redshift, $z$, and a given rest-frame bandpass with rest-frame central wavelength, $\lambda_r$, we compute a weight, $w_i$, for each observed bandpass with observed-frame central wavelength $\lambda_{o,i}$ 

\begin{eqnarray}
\gamma_i & = & \exp{(x_i^2/2/\log(1+\Delta)^2)} \\
w_i & = & W \cdot \left(\frac{2}{1+\gamma_i/\mathrm{Max}(\gamma_i)}-1\right), \label{eq:rfweight}
\end{eqnarray}

\noindent where $x_i = \log(\lambda_r) - \log(\lambda_{o,i}/(1+z))$, and $\Delta$ and $W$ are each parameters with a default value of 0.5.  That is, the weights prioritize (i.e., are smallest for) observed bandpasses that are closest to the redshifted rest-frame bandpass.  The templates are then refit to the observed photometry with least squares weights $\epsilon_i^2 = \sigma_i^2 + (w_i \cdot F_i)^2$, where $F_i$ and $\sigma_i$ are the original photometric measurements and their uncertainties, respectively.  The adopted rest-frame flux density is that of the template in this reweighted fit integrated through the target bandpass.  In short, this approach uses the templates as a guide for a weighted interpolation of observations that best constrain a targeted rest-frame bandpass that fully accounts for all bandpass shapes and relative depths of a multi-band photometric catalog.  Because the weights are localized for each target rest-frame bandpass, there is no implicit requirement that, e.g., the inferred rest-frame $V-J$ color is within the range spanned by the templates themselves.  Finally, we note that this is essentially the same methodology as that in version 1.0 of the original \texttt{eazy} code\footnote{\url{https://github.com/gbrammer/eazy-photoz/tree/8be1c9}} and now implemented in \ezp. 

\section{Star-galaxy separation}
\label{sgs}

\begin{figure}
    \centering
    \includegraphics[scale=0.45]{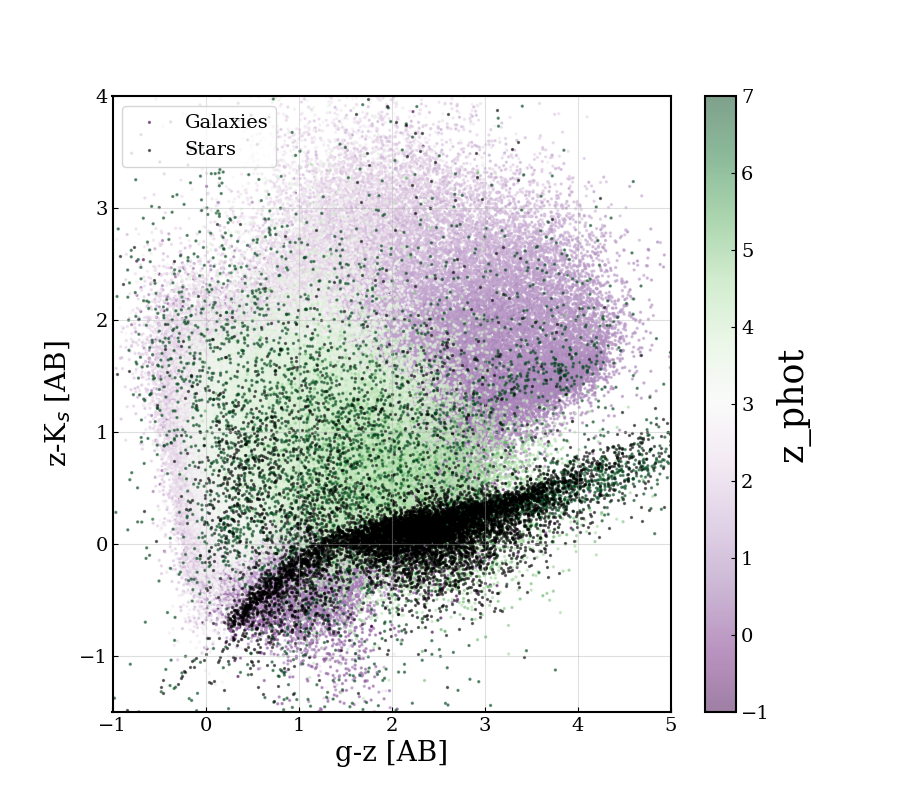}
    \caption{$gzK$ diagram for all galaxies in the combined UVISTA and HSC area with $g$, $z$ and $K$ SNR $>3$. Stars (black dots) occupy a locus across the lower part of the diagram, whereas galaxies (shown in green and purple dots) tend to occupy a wider spread of colours.}
    \label{fig:sss}
\end{figure}

\noindent We use the same method for star-galaxy separation that is described in \cite{weaver_cosmos2020_2021}. Briefly, the catalogue is matched to the Hubble ACS morphology catalogue \citep{leauthaud_weak_2007} to isolate galaxies in the half-light radii versus magnitude plane. We fit each object with stellar templates from the PEGASUS library and compute the $\chi^{2}_{star}$.  We calculate the reduced chi-squared $\chi^{2}_{r}$, where ($\chi^{2}_{r}$ = $\chi^{2}/N_{filt}-1$), of both the galaxy best fit template  and star best template where $N_{filt}$ is the number of filters used. Objects with HSC $i$-band magnitude $<21.5$ or \textit{HST}/ACS F814W magnitude $<23$ with $K_{s}$ SNR $>3$ or IRAC Channel~1 SNR $>3$ are classified as stars if they lie on the point like sequence in the half-light radii versus magnitude plane or if they are fit better with a stellar template than a galaxy template (i.e, $\chi^{2}_{star} < \chi^{2}_{galaxy}$). Additionally, objects that do not satisfy these criteria can be classified as a star if  $\chi^{2}_{star} + 1 < \chi^{2}_{galaxy}$. We find a similar fraction of stars to the number reported in the official COSMOS2020 catalogue and find a similar distribution in the $gzK$ colour colour diagram (see Figure \ref{fig:sss}).

\section{Robustness of rest-frame $J$ band estimate}
\label{app:robustness_jband}

\noindent Although searches for massive QGs at $z>3$ in the coming era will likely begin with rest-frame colour colour selections, we should be critical about which rest-frame colours should be used. The choice of rest-frame colours should depend on both the desired redshift range probed, and the ability of models to adequately measure rest-frame flux densities from the data. In particular, the colour used to discriminate between red dusty star forming galaxies and red QGs should be carefully chosen. \cite{antwi-danso_beyond_2022} explored the effects of rest-frame $J$ extrapolation for a flux limited sample of galaxies at $0.5<z<6$ and find that the flux is affected significantly at $z>2$ when extrapolation occurs. This can result in the removal of QGs in standard selections and high contamination rates. 

\begin{figure}
    \centering
    \includegraphics[width=0.45\textwidth]{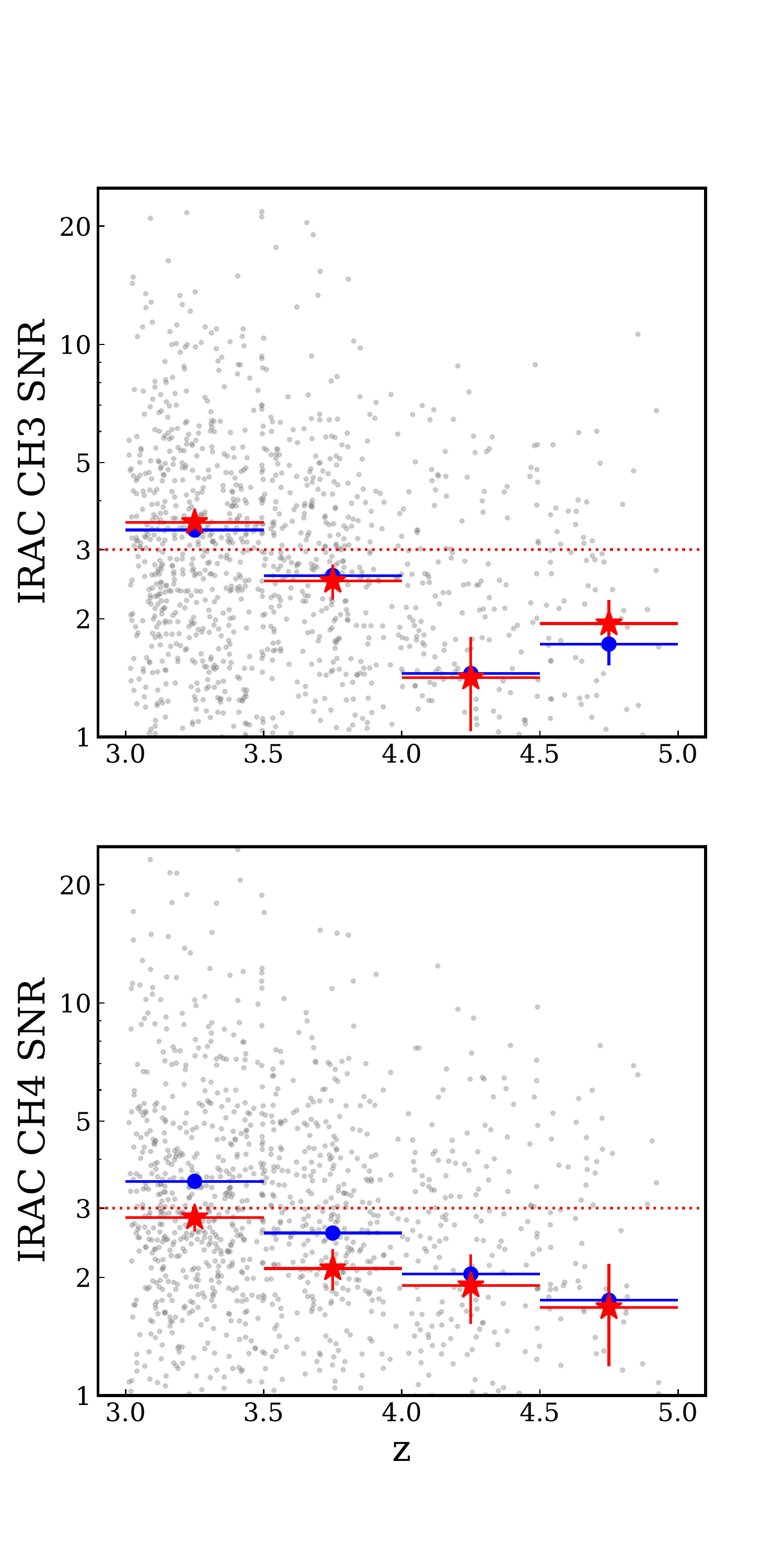}
    \caption{Signal to Noise ratio (SNR) as a function of redshift at $3.0<z<5.0$ for IRAC Channel 3 (top) and IRAC Channel 4 (bottom) for the whole massive sample (gray). The average SNR for each channel is shown binned in redshift for the entire massive sample (blue dots) and the extended quiescent sample (red stars). The worst performance is at $4.5<z<5.0$ for both channels.}
    \label{fig:irac_snr}
\end{figure}

At $z>3$, current ground-based surveys such as $COSMOS$ rely on data from $Spitzer$/$IRAC$ Channels 3 and 4. By $z\sim5$, the rest-frame $J$-band is constrained entirely by Channel 4, which has a moderately shallow limiting $3\sigma$ depth of $23.1$ mag in $COSMOS2020$. By comparison, $Spitzer$/IRAC Channels 1 and 2 are much deeper limited to 26.4 and 26.3 mags $3\sigma$, respectively. However, the fraction of galaxies in $COSMOS2020$ with $SNR\gtrsim3$ in either Channel 3 or 4 at $3<z<5$ is only $\sim5$\%.  Making progress towards understanding the physical properties of the first quenched galaxies therefore mandates deeper data in the NIR, else inventive new methods, such as the use of a synthetic rest-frame $i_{s}$ instead of $J$ \citep{antwi-danso_beyond_2022}. To ensure that rest-frame $J$ is constrained for our sample, we limited the redshift upper bound at $z<5$. Here, we compute the average SNR in bins of redshift for both Channels 3 and 4 for  the entire massive sample and the quiescent sample (see Figure \ref{fig:irac_snr}). At $3<z<3.5$ and for both bands, the mean SNR for QGs is $\gtrsim3$. At $3.5<z<4.0$, the mean SNR in Channels 3 and 4 is $2-3$, so rest-frame $J$ is still constrained by reasonable photometry. At $z>4$ however, both Channels have mean SNR between $1-2$, even after the $5\times$ boosting. This highlights the need for deeper observations in the NIR at $>5\,\mu{\rm m}$. For part of the $COSMOS$ field, this can be achieved with \textit{JWST}/MIRI within the $COSMOS-Web$ (PI: C. Casey, \citealp{casey_cosmos-web_2022}) and $PRIMER$ (PI: J. Dunlop) programs.

The issue of galaxies being classified incorrectly due to their rest-frame colours can partly be alleviated by our method of assigning a probability of quiescence to each galaxy, instead of a binary separator, which relies on the relative scatter within each population to be low -- evidently not the case at $z>3$. 

\section{Performance of an extended $UVJ$ color selection}
\label{app:extended_uvj}
At high redshift ($z\geq2$), it may make sense to lower, or remove, the $U-V>1.3$ boundary, allowing for the selection of ``recently quenched" or post-starburst galaxies. This has been suggested and implemented by multiple works. For example, \cite{schreiber_near_2018} suggested combining the lowering of the boundary with the use of a band measuring the contribution from more recent star formation, whilst \cite{marsan_number_2020}, who studied ultra-massive galaxies in the COSMOS-Ultra-VISTA field, remove the line altogether. \citealp{ belli_mosfire_2019}, \citealp{forrest_massive_2020}, \citealp{carnall_timing_2020} and \citealp{park_rapid_2022} also advocated for either removing or lowering the boundary. If the $U-V>1.3$ constraint is entirely removed, the picture changes for $UVJ$. At $z=2$ the results remain roughly the same: the TPR and FPR both increase by $4\%$. At $z>3$, the improvement is marked: the TPR increases by $25\%$ to $98\%$, whilst the FPR only increases by $3\%$ by $15\%$. $z=4$ sees the most improvement, with the TPR increasing from $34\%$ to $97\%$ whilst the FPR still remains low at $6\%$. Considering instead the contamination, which is defined as the number of galaxies defined as quiescent compared to the total number of UVJ quiescent galaxies, the conclusion changes. The contamination fraction of UVJ in \textsc{SHARK} at $z=2,3,4$ is $81,86,87\%$, which is similar to the contamination including the border ($80,86,91\%$). Therefore, although the TPR and FPR imply the extended $UVJ$ selection is suitable for use, the contamination fraction suggests otherwise, which is the same conclusion as for the classical $UVJ$ selection.

\bibliography{references}{}
\bibliographystyle{aasjournal}

\end{document}